\documentclass[aps, pra, superscriptaddress, reprint, twocolumn,floatfix]{revtex4-1}
\usepackage[usenames,dvipsnames]{xcolor}
\usepackage{appendix}
\usepackage{ulem}
\usepackage{amssymb}
\usepackage{graphicx}
\usepackage{amsmath}
\usepackage{hyperref}
\hypersetup{
 colorlinks=true,
 linkcolor=blue,
 anchorcolor = blue,
 citecolor = blue,
 filecolor = blue,
 urlcolor = blue
}
\usepackage{braket}
\usepackage{subcaption}
\usepackage{bbm}
\usepackage{dcolumn}
\usepackage{braket}
\usepackage{bm}
\usepackage{amsmath,amsthm,amssymb}
\usepackage{color}
\usepackage{verbatim}
\usepackage{comment}
\usepackage{dsfont}
\usepackage{physics}\usepackage[justification=Justified]{caption}
\begin{document}

\title{Measuring full counting statistics in a trapped-ion quantum simulator}

\newcommand{\IQOQI}{\affiliation{Institute for Quantum Optics and Quantum Information, Austrian Academy of Sciences, Technikerstra{\ss}e 21a, 6020 Innsbruck, Austria}}
\newcommand{\UIBK}{\affiliation{University of Innsbruck, Institute for Experimental Physics,  Technikerstra{\ss}e 25, 6020 Innsbruck, Austria}}
\newcommand{\SISSA}{\affiliation{SISSA and INFN, via Bonomea 265, 34136 Trieste, Italy }}
\newcommand{\ICTP}{\affiliation{International Centre for Theoretical Physics (ICTP), Strada Costiera 11, 34151 Trieste, Italy}}
\newcommand{\s}{\mathbf{s}}
\renewcommand{\tr}{\mathrm{Tr}}
\author{Lata Kh Joshi}
\SISSA
\author{Filiberto Ares}
\SISSA
\author{Manoj K. Joshi}
\IQOQI
\UIBK
\author{Christian F. Roos}
\IQOQI
\UIBK
\author{Pasquale Calabrese}
\SISSA
\ICTP

\begin{abstract}
In quantum mechanics, the probability distribution function (PDF) and full
counting statistics (FCS) play a fundamental role in characterizing the fluctuations of quantum observables, as they encode the complete information about these fluctuations. 
In this letter, we measure these two quantities in a trapped-ion quantum simulator for the transverse and longitudinal magnetization within a subsystem. We utilize the toolbox of classical shadows to postprocess the measurements performed in random bases. The measurement scheme efficiently allows access to the FCS and PDF of all possible operators on desired choices of subsystems of an extended quantum system. 
\end{abstract}
\maketitle

\paragraph*{Introduction--} One of the most striking and fundamental features of quantum mechanics is that the result of a measurement is inherently random. The possible measurement outcomes of an observable are characterized by their Probability Distribution Function (PDF) or, alternatively, by the Full Counting Statistics (FCS), the cumulant generating function of the PDF. These quantities encode all the information about the quantum fluctuations of an observable.  Understanding such fluctuations is not only essential at a fundamental level but also finds important applications in, e.g., the development of optical and electrical devices~\cite{blanter-00, gardiner, nazarov}, atomic and particle physics~\cite{Lamb_1947, QCD_Bellwied}, random number generation~\cite{QRNG_RevModPhys}, and materials science~\cite{Casimir}, to name a few.

In many-body quantum systems, there is a growing interest in the study of PDFs and FCS. The progress in developing quantum simulators have enabled accessing them experimentally~\cite{hofferberth-08, armijo-10, jacqmin-11, gring-12, bohnet-16, rispoli-19}.  These experiments consider either local observables, defined on a single point, or global ones, with support on the entire system. On the other hand, recent theoretical effort has focused on observables defined on extended subsystems. In this case, the FCS not only exhibits universal signatures both in~\cite{cherng-07, bortz-07, abraham-07, lamacraft-08, ivanov-13, eisler-13, klich-14, moreno-16, najafi-17, stephan-17, collura-17, bastianello-18, bastianello-18-2, arzamasovs-19, najafi-20, calabrese-20, ares-21, cai-24} and out-of \cite{eisler-13-2, lovas-17, groha-18, collura-19, gamayun-19, valencia-tortora-20, bertini-23, senese-23, tirrito-23, bertini-24, horvath-24, horvath-24-2, ranabhat-24} equilibrium, where it can be used as a probe of relaxation and thermalization, but is also closely related to other key quantities such as entanglement entropies~\cite{klich-09, klich-09-2, song-11, song-12, calabrese-12, levine-12, susstrunk-12, calabrese-15, utsumi-19}. 
In fact, the FCS is the simplest example of a broader class of specialized partition functions known as charged moments. These charged moments allow one to resolve the entanglement by decomposing it across the different symmetry sectors of the system~\cite{goldstein-18, xavier-18, lukin-19} and to quantify the degree a symmetry is broken in a subsystem~\cite{ares-23, rylands-24}. Although there is a large number of theoretical works, only very recently the non-equilibrium particle number PDF in an extended spatial region has been experimentally probed in a quantum gas microscope~\cite{Wienand2024} and in superconducting qubits~\cite{google-24}. In particular, the latter work employs the FCS to falsify some conjectures on the dynamical universality of Heisenberg spin chains, highlighting the relevance of studying quantum fluctuations beyond the first fewer cumulants, see also~\cite{wei-22, valli-24}. 

In this work, we measure time-evolved PDFs and FCS after a sudden quench in a trapped-ion quantum simulator, reanalyzing randomized measurement data from  Refs.~\cite{Brydges2019} and \cite{joshi-24}. We consider as observables the magnetization on a block of ions both along the transverse ($z$) and the longitudinal ($x$) axes. While the former magnetization is globally conserved by our simulated dynamics, it is not the case in the latter one. The theoretical calculation of the subsystem FCS of a non-conserved global observable is a challenging task, even in the simplest and most paradigmatic situations~\cite{lamacraft-08, collura-17, groha-18, collura-19, valencia-tortora-20}. This is one of the main reasons why previous experimental studies, as well as the majority of theoretical investigations, have been limited to globally conserved quantities. 
The randomized measurement protocol instead allows us to access the FCS and the PDF of any subsystem observable, whether or not it is globally conserved.

\begin{figure*}[ht!]
\includegraphics[width=\linewidth]{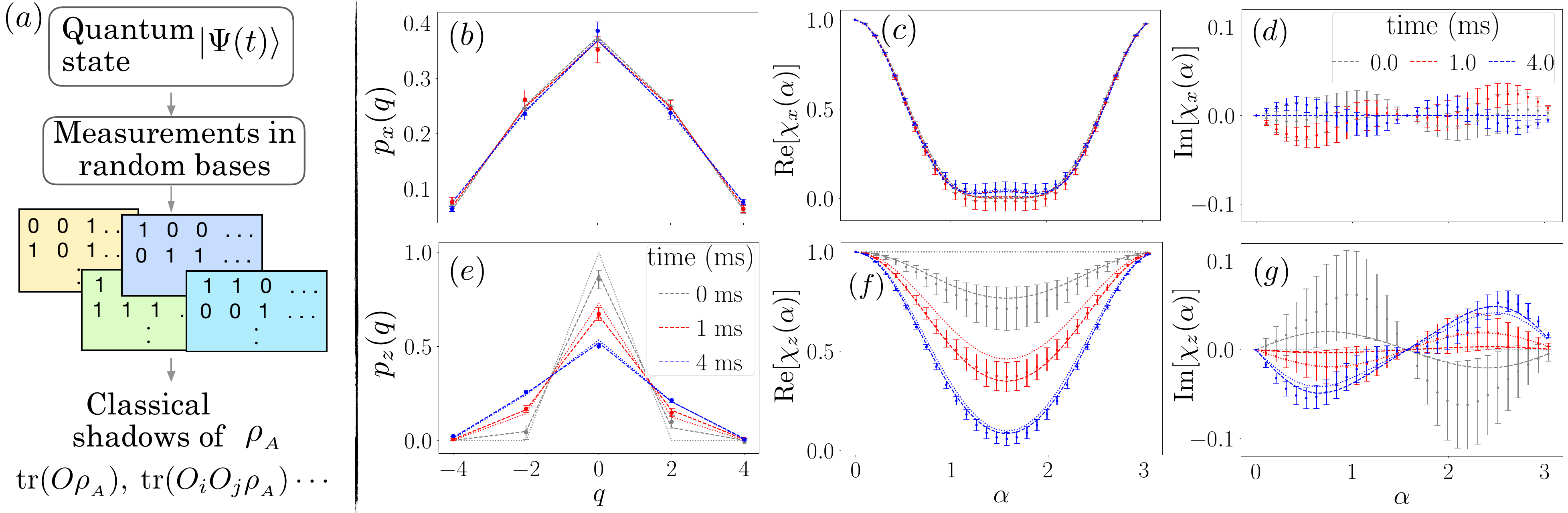}
\caption{a) \textit{Measurement protocol using classical shadows:} The protocol used to measure the full counting statistics (FCS) and the probability distribution function (PDF) of an observable $O$ in a subsystem $A$ is based on collecting randomized shadows of the reduced density matrix $\rho_A$. The quantum system is prepared in the desired state $\ket{\Psi(t)}$ and measured in Haar random bases. Experimental repetitions with various random bases provide access to the FCS and PDF.  (b, e) \textit{Magnetization PDFs in a quench from the N\'eel state:} The PDF $p_x(q)$, of the longitudinal magnetization $S_A^x$, and the PDF  $p_z(q)$, of the transverse magnetization $S_A^z$ at different times is shown. (c, d, f, g) \textit{Magnetization FCS in a quench from the N\'eel state:} Real and imaginary parts of the FCS of $S_A^x$ (in (c-d)) and that of $S_A^z$ (in (f-g))  are shown. As a proof of principle, the experiment captures the expected behavior for the PDF and FCS (see main text for details). 
In all the panels, different colors correspond to three different times; $t=0$ms (gray), $t=1$ms (red), and $t=4$ms (blue).  The dotted lines show the theoretical predictions for a perfect N\'eel state whereas the dashed lines are the predictions for initial state in presence of experimental errors. Symbols and error bars show experimental data. The total system size is $N=10$ with subsystem, of size $N_A=4$, chosen from the central sites.}
\label{fig:fig1}
\end{figure*}
\paragraph*{Basic definitions}-- Let us consider an extended quantum system in a state $\ket{\Psi}$ and divide it into two spatial parts $A$ and $\bar{A}$. We take a quantum observable $O$ which can be decomposed into the contributions from $A$ and $\bar{A}$ as $O=O_A+O_{\bar{A}}$. The FCS of $O_A$ is
\begin{equation}\label{eq:fcs}
\chi(\alpha)=\Tr(\rho_{_A} e^{i\alpha O_A})~,
\end{equation}
where $\rho_{_A}=\Tr_{\bar{A}}(\ket{\Psi}\bra{\Psi})$ is the reduced density matrix of the subsystem $A$. The derivatives of~\eqref{eq:fcs} with respect to the spectral parameter $\alpha \in[0,2\pi]$ at $\alpha=0$ give the moments of $O_A$. If $\Pi_q$ is the projector onto the eigenspace associated with the eigenvalue $q$ of $O_A$, then the PDF $p(q)$ that gives the probability of obtaining $q$ as outcome in a measurement of $O_A$ can be computed from
\begin{equation}\label{eq:pdf}
p(q)=\Tr(\Pi_q \rho_A)~.
\end{equation}
The FCS is in general a complex number. The imaginary part of $\chi(\alpha)$ probes how  asymmetric the PDF is with respect to $q=0$: if Im$[\chi(\alpha)]$=0, then $p(q)=p(-q)$. Furthermore, its slope at $\alpha=0$, ${\rm d}\mathrm{Im}[\chi(\alpha)]/{\rm d}\alpha\lvert_{\alpha=0}, $ is the mean of the PDF, i.e., $ \langle O_A\rangle$. The curvature of $\mathrm{Re}[\chi(\alpha)]$ at $\alpha=0$ gives $\langle O_A^2 \rangle$, whereas the variance of the PDF, i.e, $\sigma^2 =\langle O_A^2 \rangle-\langle O_A \rangle^2$, can be directly obtained from the logarithm of the FCS (more details in Appendix section A).

Our experimental protocol builds on the versatile randomized measurement toolbox \cite{Huang2020,elben_review}. The quantum system of $N$ qubits is prepared in a particular configuration $\ket{\Psi_0}$ and quenched with the engineered Hamiltonian $H$ to reach the time evolved state $\ket{\Psi(t)}$, see, e.g., Refs.~\cite{Brydges2019, joshi-24}. As summarized in Fig.~\ref{fig:fig1}~(a), on this state measurements are performed in random bases. For each random basis, the experiment is repeated to collect quantum projections. These data are sufficient to obtain measurements of FCS and PDF (see Appendix section B for details).  Randomized measurements have been useful to access fundamental non-equilibrium properties of quantum systems, from entanglement entropies in pure~\cite{Brydges2019,Rath2021,Satzinger2021,Hoke2023, rath-23} and mixed states~\cite{Zhou2020,Elben2020b,Neven2021}, rate of information scrambling~\cite{Joshi2020a} to quantum chaos~\cite{LKJ_2022_chaos, dong2024measuring}. 
In this letter, we use the randomized measurement data from Refs.~\cite{Brydges2019, joshi-24}. The key new element in our work is the post-processing of the experimental data to access the FCS and PDFs in these experiments. 

\paragraph*{Setup--}
The experiment we use is a quantum simulator based on trapped Ca$^+$ ions.  The qubit basis states, labeled by the vectors $s = (\ket{\downarrow}, \ket{\uparrow})\equiv (1,0)$, are encoded into the $\ket{\text{S}_{1/2}, m=+1/2}$ and $\ket{\text{D}_{5/2}, m=+5/2}$ electronic levels of the ions. The engineered power-law decaying  Hamiltonian is
\begin{equation}
H_{\mathrm{XY}} = \sum_{i>j}\frac{J_0}{2|i-j|^\alpha} \left(\sigma^x_i \sigma^x_j + \sigma^y_i \sigma^y_j\right)~,
\label{eq:IsingH}
\end{equation}
where $\sigma^a_i$ are the spin-$1/2$ Pauli matrices for $a=x,y$ at the lattice site $i=1\dots N$. We study two cases from two independent runs of the experiment: \textbf{I)} In the first case, as in Ref.~\cite{Brydges2019}, the initial state is a N\'eel state, $\ket{\Psi_0}=\ket{\uparrow\downarrow\uparrow\downarrow\dots}$, in a total system of $N=10$ spins. The interaction strength $(J_0)$ and range $(\alpha)$ in the quenching Hamiltonian are $J_0 \approx 420~\mathrm{rad/s}$ and  $\alpha \approx 1.24$; \textbf{II)} In the second case, as in Ref.~\cite{joshi-24}, the initial state is a tilted ferromagnet: a ferromagnetic state $\ket{\downarrow\downarrow\downarrow\dots}$ rotated by an angle $\theta$ away from the $z-$axis, 
\begin{equation}
|\Psi_0(\theta)\rangle=e^{i \frac{\theta}{2}\sum_j\sigma^y_j}|\downarrow\downarrow\downarrow\dots\rangle~.
\label{eq:tiltedFerro}
\end{equation}
The total system size studied is $N=12$ spins with $J_0 \approx 560~\mathrm{rad/s}$ and  $\alpha \approx 1$. 

For the purpose of illustration, we measure the FCS $\chi_\mu(\alpha)$ and PDF $p_\mu(q)$ of $S^\mu_A=\sum_{j\in A}\sigma_j^\mu$, which are the transverse ($\mu=z$) and longitudinal ($\mu=x$) magnetization in a subsystem $A$ 
\footnote{Note that we have defined $S_A^z$ and $S_A^x$ as twice the usual magnetizations in order that their eigenvalues are integers.}.  
The full system transverse magnetization is conserved by the post-quench Hamiltonian, i.e. $[S_{A+\bar A}^z, H_{\rm XY}]=0$, but the longitudinal one is not, $[S_{A+\bar A}^x, H_{\rm XY}]\neq 0$. The FCS of both magnetizations satisfies $\chi_\mu(\alpha+k\pi)=(-1)^{kN_A}\chi_\mu(\alpha)$ where $k\in \mathbb{Z}$ and $N_A$ is the number of spins in $A$. Therefore, $\chi_{\mu}(\alpha)$ is a periodic function of period $\pi$ ($2\pi$) for even (odd) $N_A$.

\paragraph*{Case I: PDF and FCS in the quench dynamics of the N\'eel state--}In Fig.~\ref{fig:fig1}~(b), we present the PDF of the longitudinal magnetization $S_A^x$ at times $t=0, 1$ and $2$ ms (in gray, red and blue) in a system of $10$ ions and subsystem of size $N_A=4$. For a N\'eel state,  $p_{x}(q)=2^{-N_A}\binom{N_A}{(N_A-q)/2}$ and we obtain a good match between the theory (dashed curve) and the experimental data (dots and error bars). Notice that, under the quench dynamics, the PDF $p_x(q)$ changes slightly and remains symmetric around the eigenvalue $q=0$. This is further confirmed by the measurements of the corresponding FCS $\chi_x(\alpha)$ presented in Fig.~\ref{fig:fig1}~(c-d). For a N\'eel state, the FCS is $\chi_x(\alpha)= \cos^{N_A}(\alpha)$. We see that the imaginary part remains zero within error bars at all times. After the quench~\eqref{eq:IsingH} of the N\'eel state, $\langle S_A^x\rangle=0$, which agrees with the zero slope of the imaginary FCS at $\alpha=0$ at all times. Interestingly, the curvature of the real part of FCS at $\alpha=0$, i.e. $\langle(S_A^x)^2 \rangle$, coincides with the variance of the PDF in the present case since $\langle S_A^x\rangle=0$. In Appendix section C, we show $\langle(S_A^x)^2\rangle$ as a function of time. 

As stated in the introduction, the same experimental dataset used to access the PDF and FCS of the subsystem magnetization along one axis can also be used to measure it along any arbitrary direction. We next consider the transverse magnetization $S_A^z$. The N\'eel state, for an even number of sites in the subsystem, is an eigenstate of $S_A^z$ with zero eigenvalue; therefore, we expect the corresponding PDF to be $p_{z}(q)=\delta_{q,0}$. Since $S_A^z$ is well defined in the initial state and is globally conserved by the post-quench Hamiltonian, it only fluctuates within subsystems during the time evolution, as seen in its PDF in Fig.~\ref{fig:fig1}~(e). In the experimental data at time $t=0$ in panel (e) (the gray points), we notice that $p_z(q=\pm 2) > 0$ whereas $p_z(q=0 )<1$. Such mismatch of the ideal theory (dotted curves) and  experimental results can be explained by considering state preparation and  measurements errors. At $t=0$, these errors appear  as random local bit-flips in the final measurements. Modeling the initial state in presence of experimental errors, in Fig.~\ref{fig:fig1}~(e), we plot the PDF $p_z(q,t)$ using dashes and observe a good match between the theory and the experiment. In SM section D,  details of the error model and estimation of the error rates are presented.

Next, in Figs.~\ref{fig:fig1}~(f-g), the FCS of $S_A^z$ is measured. At $t=0$ for even sized subsystems, we expect $\chi_z(\alpha)=1$ (dotted gray line). In the measured FCS, however, we see a different behavior, shown in the figure with symbols and error bars. By introducing effective errors in the theoretical state, we find that the experimental data is well explained with the theory (dashed curves) (see SM section D for details of the error model). We can see that the effect of experimental errors becomes less relevant with time, and the difference between the theoretical FCS and the measured FCS reduces. Also note that as time passes the curvature of the real part of the FCS increases, which is reflected in an increasingly wider PDF, whereas the very small imaginary part of FCS corresponds to a near symmetric PDF at all times. The behavior of the FCS with time is presented in Appendix SM section E. The error bars shown are the standard error of the mean over finite random unitaries. In Appendix section F we detail the error bars and their propagation.

\begin{figure}[t]
\includegraphics[width=\linewidth]{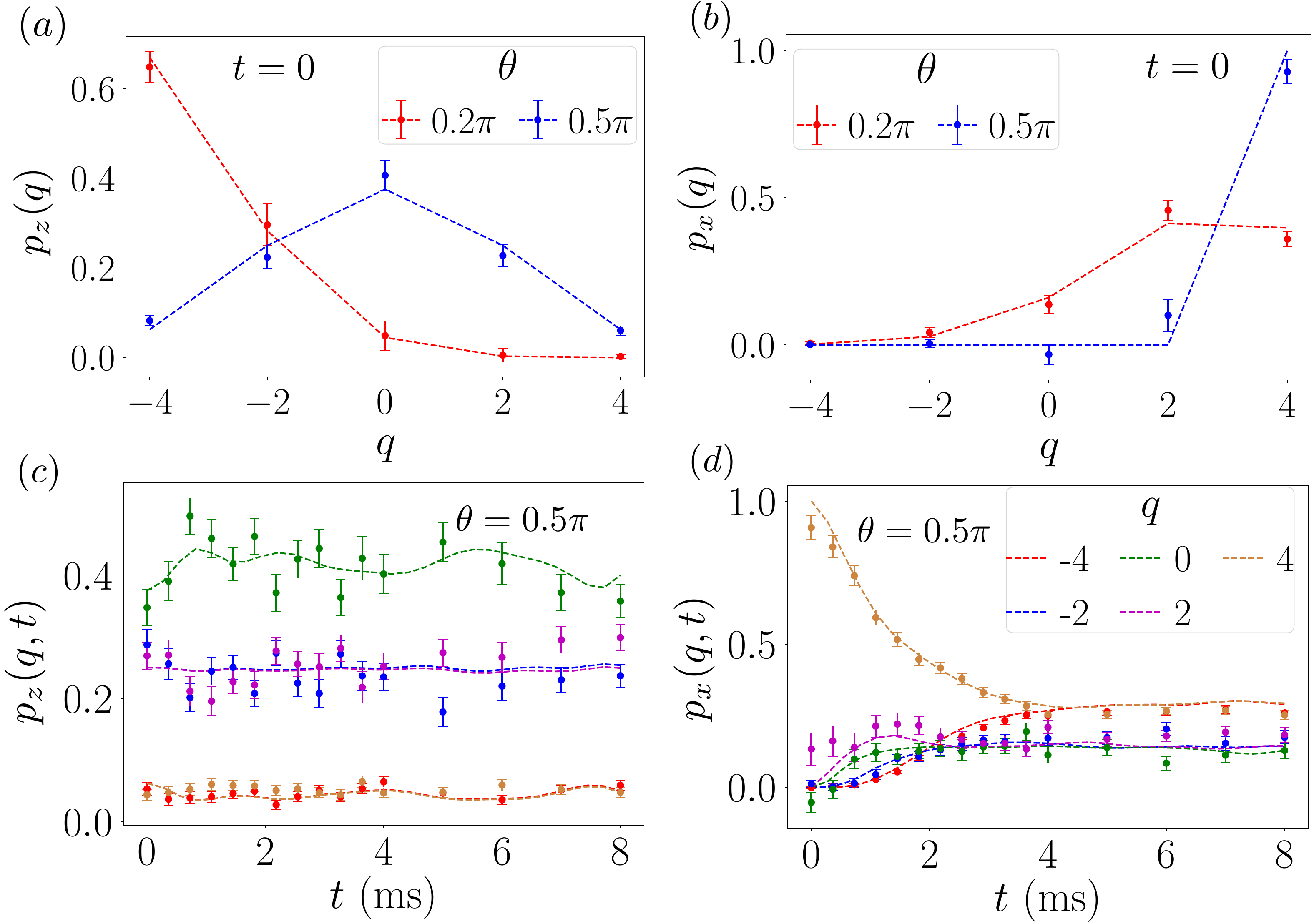}
\caption{\textit{Magnetization PDF in the quench from tilted ferromagnets:} The PDF in a system of $N=12$ and $N_A=4$, for two of the magnetizations, $S_A^z$ (panels (a-c)) and $S_A^x$ (panels (b-d)) are shown. Dashed curves (theory including decoherences during the quench dynamics) match the experimental data (dots and error bars). (a-b) The PDFs are shown as functions of the possible measurement outcomes $q$ for the tilt angles $\theta=0.2\pi$ and $0.5\pi$ at $t=0$. (c-d) We plot the PDFs during the quench dynamics for several values of $q$ for initial state with $\theta=0.5\pi$. }
\label{fig:fig3}
\end{figure}

\paragraph*{Case II: PDF and FCS in the quench dynamics of a tilted ferromagnetic state--} We now consider the other class of initial states; namely, the tilted ferromagnets~\eqref{eq:tiltedFerro}.  These states were realized in the recent experiment \cite{joshi-24},
on a system of $N=12$ qubits. These states make an interesting example because their PDFs are distinct from that of a N\'eel state. In Fig.~\ref{fig:fig3}~(a), we present the outcome probabilities in the measurement of $S_A^z$ in a subsystem of $N_A=4$, and two choices of the rotation angle, $\theta=0.2\pi$ and $0.5 \pi$, at time $t=0$.   For $\theta=0.5\pi$, we get $p_z(q, \theta=0.5\pi)=2^{-N_A}\binom{N_A}{(N_A-q)/2}$, the blue curve. 
The red curve is the theoretical PDF $p_z(q)$ for the angle $\theta=0.2\pi$. In Fig.~\ref{fig:fig3}~(b), we show the PDF for the longitudinal magnetization $S_A^x$ at $t=0$. For $\theta=0.5\pi$, we expect $p_x(q)=\delta_{N_A-q,0}$ (blue dashes), whereas we notice some deviation in the experimental data. This can be both an artifact of the choice of the subsystem and an effect of the measurement errors (see Appendix section G).
As the tilting angle $\theta$ decreases, the PDF $p_x(q)$ becomes wider and its peak is displaced towards $q=0$, as happens for $\theta=0.2\pi$ (dashed red curve).

In Figs.~\ref{fig:fig3}~(c-d), we show the time evolution of the PDF for $S_A^z$ and $S_A^x$, respectively, after the quench from a tilted ferromagnet with $\theta=0.5\pi$. Their dynamics show different behavior as discussed in the following.  While for the $z$ component, $p_z(q)$ is an even function in $q$ and remains almost constant in time for all values of $q$, this is not the case in the $x$ magnetization. At $t=0$, $p_x(q)$ is non-zero for $q=4$ only (panel (d)). After the quench, $p_x(q=4)$ decreases until it becomes stationary at late time, around $t\approx 4$ ms. On the other hand, the probabilities of the rest of the allowed values of $q$ increase under the time evolution. The evolution with time happens such that, at large times, $p_x(q)$, which was initially a delta function, becomes symmetric under $q\mapsto -q$. All this phenomenology is well captured by the experimental data. In the theoretical curves, in addition to the Hamiltonian evolution, decoherences during time-evolution are taken into account \footnote{ The decoherences and their rates have been taken from  Ref.~\cite{Brydges2019} and \cite{joshi-24}}.

\begin{figure}[t]
\includegraphics[width=\linewidth]{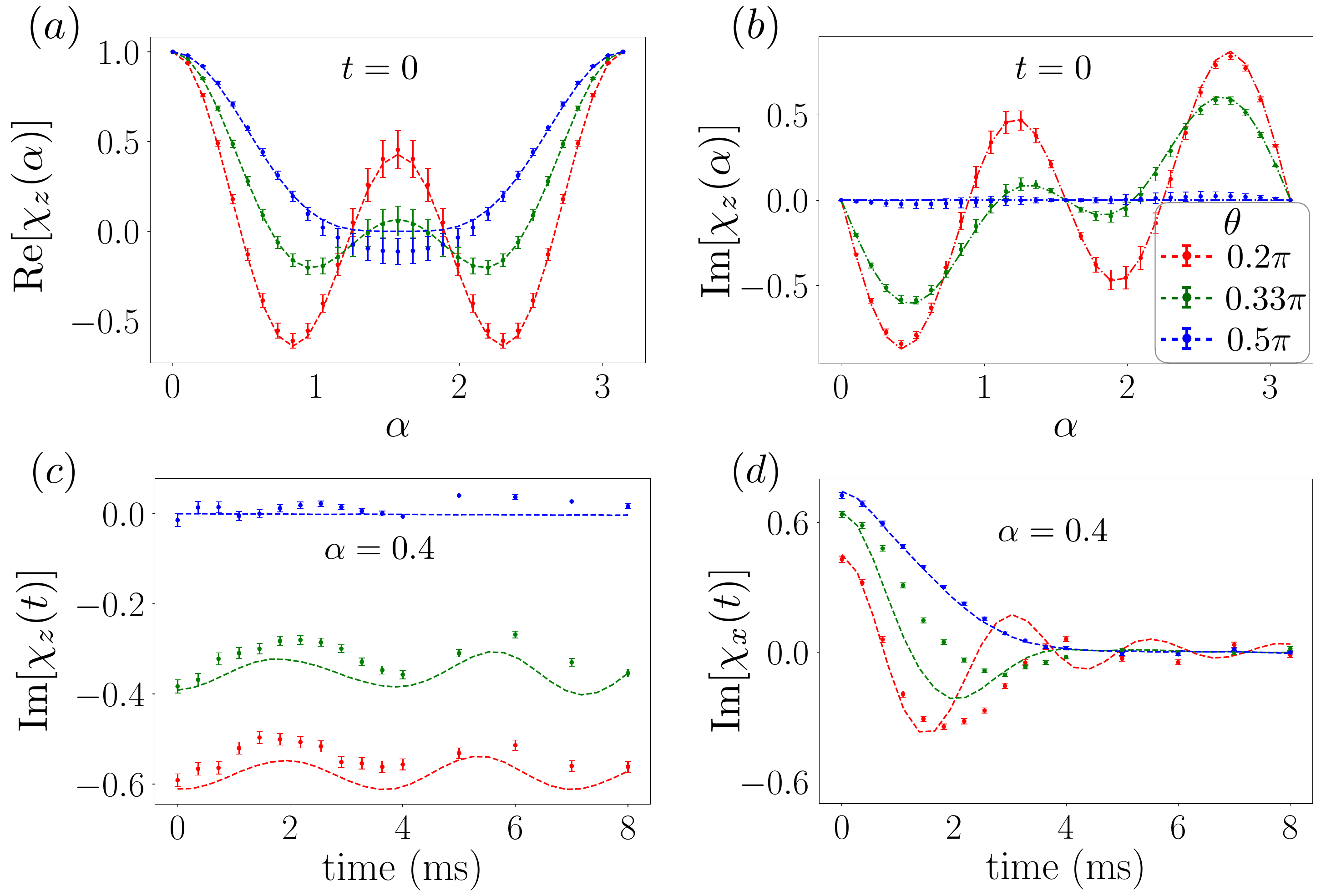}
\caption{\textit{Magnetization FCS in a quench from tilted ferromagnets:} FCS in a system of $N=12$ sites and subsystem of size $N_A=4$ is shown for different initial states. A good match between theory (dashed curves) and experiment (dots and error bars) is observed. (a-b) Real and imaginary parts of the FCS $\chi_z(\alpha)$ for the $S_A^z$ at initial time $t=0$ as functions of the spectral parameter $\alpha$. We see different behavior for different tilt angles $\theta$ (red- $0.2\pi$, green- $0.33\pi$, blue- $0.5\pi$). (c-d) Imaginary part of the FCS $\chi_\mu (t)$ for $\mu=z$ and $x$ under the quench \eqref{eq:IsingH}. The behavior of FCS with $\alpha$ and $t$ is consistent with the PDFs reported in Fig.~\ref{fig:fig3} (see main text).
}
\label{fig:fig4}
\end{figure}
In connection to the measurements of the magnetization PDF in tilted ferromagnets, in Fig.~\ref{fig:fig4} we show the corresponding FCS $\chi_\mu(\alpha)$ for three values of the initial tilt angle, $\theta=0.2\pi$ (in red), $0.33\pi$ (in green), and $0.5\pi$ (in blue). The experimental data points follow theory predictions (dashed curves) thus providing a good estimation of the FCS. 

For the states $\ket{\Psi_0(\theta)}$ \eqref{eq:tiltedFerro}, the FCS $\chi_z(\alpha)=(\cos\alpha-i\sin\alpha\cos\theta)^{N_A}$, and $\chi_x(\alpha)=(\cos\alpha+i\sin\alpha\sin\theta)^{N_A}$. 
In Figs.~\ref{fig:fig4}~(a-b), we present measurements of the real and imaginary parts of $\chi_z(\alpha)$. The imaginary FCS, near $\alpha=0$, is smallest for the largest of the three angles and it increases in magnitude with the decreasing angle. This observation is consistent with the symmetry of the PDF in Fig.~\ref{fig:fig3}~(a), i.e., the PDF is symmetric (asymmetric) for the angle $0.5\pi$ ($0.2\pi$). Moreover, the negative slope of the imaginary FCS for $\theta=0.2\pi$ at $\alpha=0$ indicates a negative value of the mean of the corresponding PDF $p_z(q)$, cf. the red curve in Fig.~\ref{fig:fig3}~(a). Another physically interesting value of the FCS is at $\alpha=\pi/2$. At this point for $\theta=0.2\pi$, the real part exhibits a peak, that becomes smaller as $\theta$ increases and disappears for $\theta=0.5\pi$. At $\alpha=\pi/2$, $\chi_\mu(\alpha)$ is the expectation value, $\chi_\mu(\pi/2)=i^{N_A}\langle P_\mu\rangle$, of the spin parity operator $P_\mu=\prod_{j\in A}\sigma_j^\mu$ in
the $\mu$ axis. For the tilted ferromagnet, $\langle P_\mu\rangle$ is the component of the spins in the $\mu$ axis; in particular, $\langle P_z\rangle=(-\cos\theta)^{N_A}$. For the behavior of $\chi_x(\alpha=\pi/2)$, see Appendix.

The symmetry of the PDF during the time evolution can further be related to the FCS. In Figs.~\ref{fig:fig4}~(c-d), the imaginary parts of the FCS under quench dynamics are shown. We see that Im$[\chi_z(t)]$ nearly remains constant with time in contrast to the time evolution of Im$[\chi_x(t)]$ which approaches zero at late times. We can compare the results for the tilt angle $0.5\pi$  with those obtained for the corresponding PDF in Figs.~\ref{fig:fig3}~(c-d). The nearly zero value of the Im$[\chi_z(t)]$ during the evolution (the blue curve in the Fig.~\ref{fig:fig4}(c)) is consistent with the symmetry $p_z(-q)=p_z(q)$ for all $t$, as in the Fig.~\ref{fig:fig3}~(c). On the other hand, the late time symmetrizing behavior of $p_x(q, t)$ in Fig.~\ref{fig:fig3}(d), is manifested in the late time relaxation of Im$[\chi_x(t)]$ to a zero value.

\paragraph*{Conclusions and outlook--} In this letter, we have presented measurements of the full counting statistics (FCS) and the probability distribution function  (PDF) of observables in subsystems of extended quantum systems prepared in a trapped-ion quantum simulator. We reanalyzed the experimental data from two previous experiments ~\cite{Brydges2019, joshi-24} where quantum quenches with a long-range Ising type Hamiltonian (Eq.~\eqref{eq:IsingH}) have been performed. The protocol allows access to the FCS and PDFs of any operator independent of them being symmetries of the evolution Hamiltonian. 
Such efficiency of the protocol makes our measurement scheme useful in characterizing the prepared quantum states along all possible magnetization directions. In particular, the comparison between the antiferromagnetic and tilted ferromagnetic states illustrates that the PDF and FCS are powerful diagnostic tools for probing and distinguishing distinct quantum phases of matter realized in quantum simulators. In our examples, we have found FCS to be highly effective for detecting experimental errors. The FCS encodes all moments of the operator, allowing us to detect errors without the need to check many observables.

Although we have focused the current demonstration on a 1D chain using a trapped-ion quantum simulator, our measurement scheme can be directly adapted to other quantum platforms where single-site controls and detections are feasible. Note that, the theory of  randomized measurements finds interesting applications in higher dimensions, as has already been demonstrated in a recent remarkable experiment studying thermalization~\cite{andersen2024thermalization}. FCS in 1D and beyond is a useful tool to distinguish quantum phases~\cite{PhysRevLett.133.083402}, probe mixed-state topology~\cite{zang-24, mao-24}, spin hydrodynamics~\cite{McCulloch2023} and studying thermalization~\cite{senese-23}. Therefore, having these protocols open up new directions especially when we lack the theoretical tools to analyze time-dependent quantum fluctuations. 

\paragraph*{Acknowledgements--} LKJ acknowledges support from the European Union’s Horizon Europe program under the Marie Sklodowska Curie Action Project ETHOQS (Grant no. 101151139). PC and FA acknowledge support ERC-AdG grant MOSE No. 101199196 and from European Union - NextGenerationEU, in the framework of the PRIN Project HIGHEST no. 2022SJCKAH\_002.  MKJ and CFR acknowledge the funding under Horizon Europe programme HORIZON-CL4-2022-QUANTUM-02-SGA via the project 101113690 (PASQuanS2.1). 

\section*{Appendices}
\subsection{Properties of the FCS and PDFs}
\label{sec:fcsandpdf}
The full counting statistics (FCS) and the probability distribution function (PDF) are related by the Fourier transform
\begin{equation}\label{eq:rel_pdf_fcs}
\chi(\alpha)=\sum_q p(q) e^{i\alpha q}.
\end{equation}
If we take the Taylor expansion of the FCS $\chi(\alpha)$ around 
$\alpha=0$,
\begin{equation}
\chi(\alpha)=1+i\alpha \langle O_A\rangle -\frac{1}{2}\alpha^2\langle O_A\rangle^2+\cdots,
\end{equation}
then we can see that its real and imaginary parts correspond to the 
even and odd powers in $\alpha$ respectively. This means that,
if $\chi(\alpha)$ is real, then it is an even function in $\alpha$ and, according to Eq.~\eqref{eq:rel_pdf_fcs}, yields a PDF $p(q)$ symmetric in $q$, i.e. $p(q)=p(-q)$. Therefore, the imaginary part of $\chi(\alpha)$ informs us about how asymmetric is the PDF with respect to $q=0$.
In particular, its leading order term in the expansion around $\alpha=0$ is the mean $\mu=\langle O_A\rangle$ of the corresponding PDF. In combination with the quadratic term of the  real part, we can also determine its variance, $\sigma^2=\langle O_A^2\rangle-\langle O_A\rangle^2$. Therefore, a larger positive 
(negative) slope of ${\rm Im}(\chi(\alpha))$ at $\alpha=0$ indicates that the PDF has a larger positive (negative) mean. When the mean is small, $\mu\approx 0$, we can approximate the variance as $\sigma^2\approx \langle Q_A^2\rangle$. In that case, a larger curvature of ${\rm Re}(\chi(\alpha))$ at $\alpha=0$ tell us that the corresponding PDF has a larger variance. 

Both $\langle O_A\rangle$ and $\langle O_A^2\rangle$ can be extracted from the randomized measurement data. Therefore, the mean $\mu$ and the variance $\sigma^2$ can in principle be determined without resorting to the knowledge of the FCS. However, the calculation of $\sigma^2$ using $\sigma^2=\langle O_A^2\rangle-\langle O_A\rangle^2$ suffers from catastrophic cancellation: a very good approximation of $\langle O_A\rangle$ and $\langle O_A^2\rangle$ may yield a very bad approximation of $\sigma^2$ if both terms are of the same order of magnitude. The FCS might provide a solution to this issue as the leading order term of its logarithm around $\alpha=0$,
\begin{equation}
\log\chi(\alpha)=i\mu\alpha-\frac{1}{2}\sigma^2\alpha^2+\cdots
\end{equation}
gives direct access to $\sigma^2$.  

In the case of the magnetization, $\alpha=0$ is not the only point 
of $\chi(\alpha)$ that provides physically relevant information. Taking into account that $e^{i\alpha \sigma_j^{\mu}}=\cos(\alpha) I+i\sin(\alpha)\sigma_j^\mu$, we have that $\alpha=\pi/2$ is the expectation value $\chi_\mu(\pi/2)=i^{N_A}\langle P_\mu\rangle$ of the spin parity operator $P_\mu=\prod_{j\in A}\sigma_j^\mu$ in the $\mu$ axis. In particular, if the subsystem $A$ is in a product state with all the spins aligned with the $\mu$ axis, then $\langle P_\mu\rangle = (-1)^{n_-^\mu}$ where $n_-^\mu$ is the number of spins pointing in the negative direction of $\mu$. In the tilted N\'eel state, since all the spins are aligned with the $z$ axis, we have that $\langle P_z\rangle=1$ and $\langle P_x\rangle=0$ for an even number of ions in $A$. This explains why in the Fig.~1 in the main text (MT), the FCS vanishes at $\alpha=\pi/2$ for the longitudinal magnetization but not for the transverse one. Observe that $\langle P_z\rangle$ is strongly affected by the bit-flip errors in the state preparation and by the time evolution, being very close to zero at $t=4$ ms. 
On the other hand, $\langle P_x\rangle$ exhibits only minor oscillations around zero throughout the entire time evolution.
In the tilted ferromagnetic state, the spins have certain component both in the $z$  and $x$ directions and, therefore, $\langle P_z\rangle=(-\cos\theta)^{N_A}$ and $\langle P_x\rangle=(\sin\theta)^{N_A}$.
\subsection{Classical shadows, measurement estimators and random unitaries (RU)}
\label{sec:SM_shadows}
We use classical shadows to estimate the FCS and PDF. 
After the state $\rho(t)$ is prepared, projective measurements are performed in random bases. This is achieved by applying local random rotations $U=\otimes_i u_i$, where $u_i$ are drawn from a unitary 2-design, to reach the state $U\rho(t) U^\dagger$. On this state, projective measurements in the computational $z-$basis are performed. For the same set of $u_i$, the experiment is repeated to collect $N_M$ projections in the $z-$basis. At each measurement, we obtain the bitstrings $s_i^{(r, m)}$, where the site index $i$ runs in the interval $i=1, \dots, N_A$ and $m=1,\dots, N_M$. The experiment is repeated for $N_u$ choices of random unitaries, collecting $N_M$ projections for each random $U$.  Thus for $r$-th set of random unitaries at a local site $i$, the measurement shots are $\mathbf{s}_i^{(r)}=\{s_i^{(r,1)},s_i^{(r,2)},\cdots, s_i^{(r,N_M)}\}$. These data sufficiently provide the shadow of the reduced density matrix for the $r$-th random unitary. 

The classical shadow $\hat{\rho}_A^{(r)}$ of the density matrix $\rho_A$ is constructed by averaging over the shadows $\hat{\rho}_A^{(r, m)}$ of the $N_M$ measured bitstrings for an applied unitary $U^{(r)}$,  $\hat{\rho}_A^{(r)} = \mathbb{E}_{N_M}[\hat{\rho}_A^{(r,m)}]$, where each of the  $\hat{\rho}_A^{(r,m)}$ is~\cite{Huang2020, elben_review} 
\begin{equation}
\hat{\rho}^{(r,m)}_{A} = \bigotimes_{i \in A} 3 \,{u_i^{(r)}}^\dag \ketbra{s_i^{(r,m)}}{s_i^{(r,m)}} u_i^{(r)} - \mathbb{I}_2~.
\end{equation}
The classical shadows form an unbiased estimator of the density matrix of interest, that is, the average over the applied unitaries and the measurement outcomes gives $\mathbb{E}[\hat{\rho}^{(r,m)}_A] = \rho_A$.

Our estimator for FCS uses the shadows $\hat{\rho}_A^{(r)}$ to obtain the estimator
\begin{equation}
\widehat{\chi} (\alpha) = \frac{1}{N_u}\sum_{r=1}^{N_u}\tr\left[ \hat{\rho}_A^{(r)} e^{i\alpha O_A}\right],
\end{equation}
such that $\mathbb{E}[\widehat{\chi} (\alpha)]={\chi} (\alpha)$. Similarly, the estimator for the corresponding PDF reads 
\begin{equation}
\widehat{p} (q) = \frac{1}{N_u}\sum_{r=1}^{N_u}\tr\left[ \hat{\rho}_A^{(r)} \Pi_q\right]
\end{equation}
such that $\mathbb{E}[\widehat{p} (q)]={p} (q)$. Note that, since the random unitaries $U^{(r)}$ are uniformly distributed, these estimators are agnostic to the spatial orientation of the observable $O_A$ considered, allowing to obtain the FCS and PDF of globally conserved as well as globally non-conserved operators with the same measurement data.  
\begin{figure}
\includegraphics[width=0.42\textwidth]{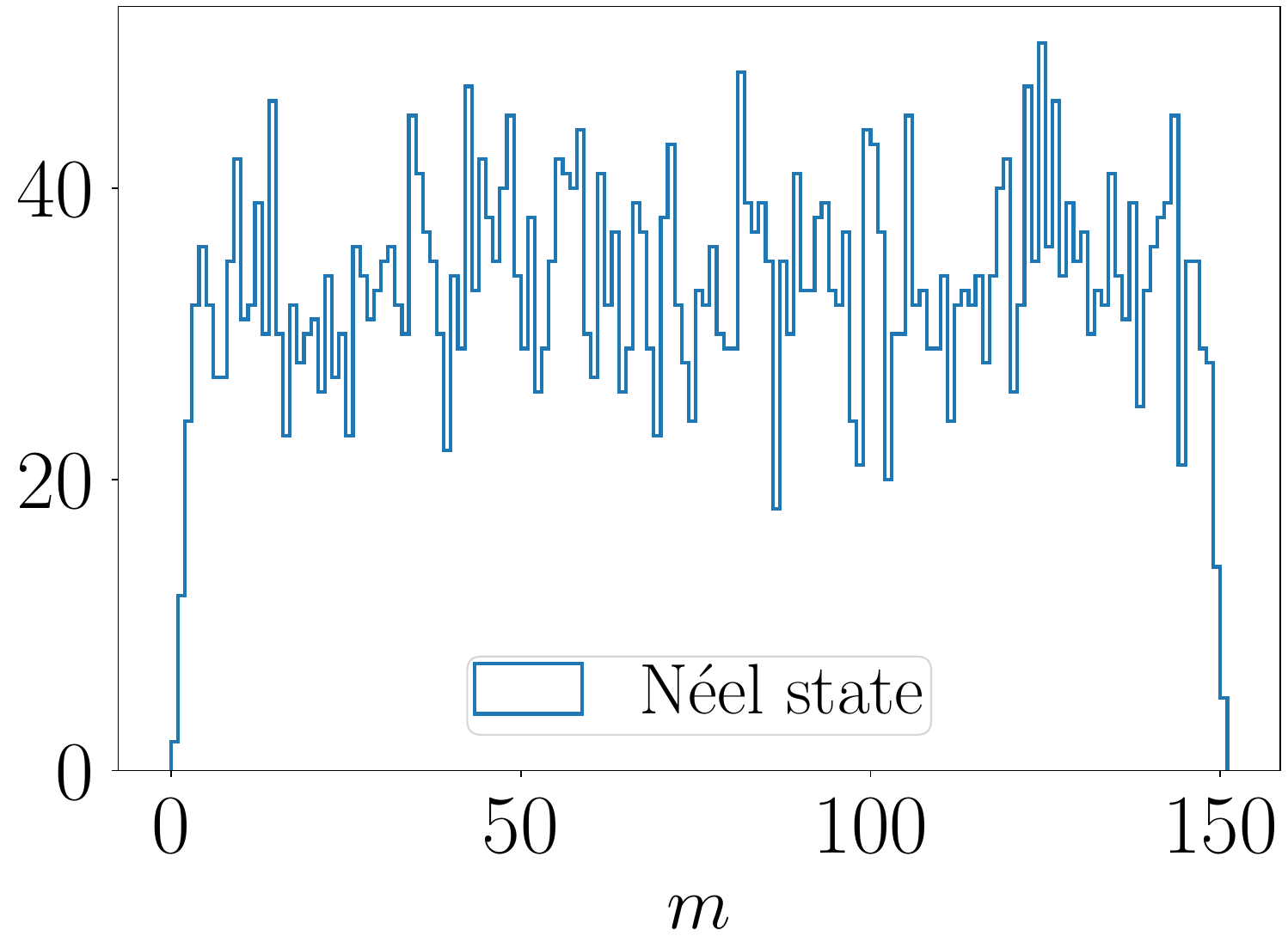}
\includegraphics[width=0.42\textwidth]{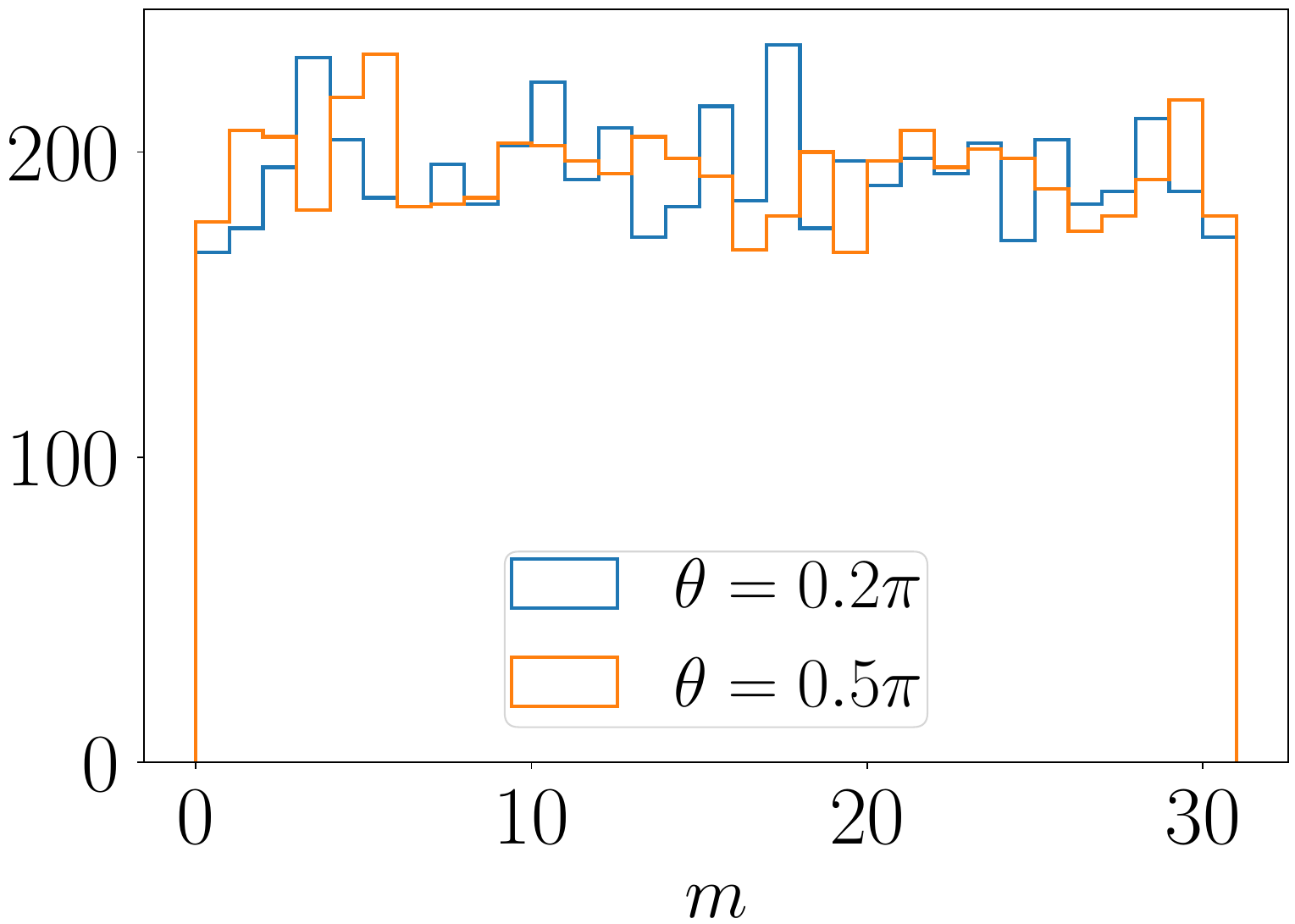}
  \caption{Histograms of spin projections following applications of random unitaries. On the $x$-axis, the label $m$ is the count of up spin measured at local sites. For a good statistics, we have combined the measurement data on all sites. (Top): uniform distribution of implemented RUs is seen in the experiment beginning with a N\'eel state. The experiment has 500 RUs with 150 shots per RU.  (Down): uniform distribution of implemented RUs is seen in the experiment beginning with tilted ferromagnetic states; blue- data for the state with tilting angle $\theta=0.2\pi$ and orange- data for the state with tilting angle $\theta=0.5\pi$. The experiment has 500 RUs with 30 shots per RU.  }
\label{fig:histograms_RU}
\end{figure}

\subsubsection{Generation of local random unitaries}
A set of single qubit random unitaries are chosen from a circular uniform distribution (CUE) \cite{Brydges2019, Mezzadri2006}. On the experimental side, a unitary $u_i$ on the $i-$th qubit is decomposed into the $X$, $Y$ and $Z$ rotations with optimized angles. The optimized decomposition is carried out to account for local single-qubit $Z$ rotation and global single-qubit $X$ and $Y$ rotations. The local single-qubit rotation about the $Z$ axis is carried out using a tightly focused offresonant laser beam implementing the AC Stark shift on the qubit levels. On the other hand, global single-qubit rotations are implemented by a global 729 nm laser beam that couples the qubit levels resonantly. The experimental sequence for the full single-qubit unitary on the whole ion chain is employed in parallel such that a minimum number of laser pulses are employed to achieve the full unitary $U =  \otimes_{i=1}^{N} u_i$. A total of 200 to 500 random unitaries are employed in the experimental platform. Further details about the implementation of random unitaries can be found in the supplementary material of Ref.~\cite{Brydges2019}. 

\subsubsection{Characterization of applied random unitaries}
Applied random unitaries are assessed by characterizing the uniformity of weight of eigenvectors of $O=\sigma_z^{\otimes N}$. To this end, we look at the probability distribution of randomized outcomes.
A correct implementation of the CUE is expected to give a uniform distribution in the Pauli basis. The characterization of the implemented RUs is done at $t=0$ where the decoherence effects are minimal. In both experiments reported in the present work, the initial states are product states $\ket{\psi}=\otimes_{i=1}^N\ket{\psi_i}$. These are applied with local RUs resulting in states, $\ket{\psi_f}=\otimes_{i=1}^N u_i\ket{\psi_i}$, which are then measured in the $z-$basis. For uniformly distributed $u_i$, we expect the resulting qubit state on each site to be uniformly distributed on the Bloch sphere. Thereby, the projective measurements over $N_u$ repetitions of random unitaries will span all possible orientations uniformly. For each RU, $N_M$ spin projections are taken. In SM Fig.~\ref{fig:histograms_RU}, we plot the distribution of the measured projections. The histograms show the number of times $m$, where $ 0\le m \le N_M$, a spin is projected in $\ket{\uparrow}\equiv0$ state. We notice uniform distributions over $N_M$ possibilities, confirming the uniform sampling of the RUs. In the N\'eel state data (SM Fig.~\ref{fig:histograms_RU} (top)), we also notice the fall off of the distribution at the edges, i.e., a fall off in observing either all 0 or all 1 over $N_M$ shots is seen. This indicates that a pure state is not observed due to experimental errors.

\subsection{Measurements of $\langle S_A^x\rangle$ and $\langle (S_A^x)^2\rangle$ in a quenched N\'eel state}
\label{sec:curvatureNeel}
\begin{figure}[h!]
\centering
\includegraphics[width=\linewidth]{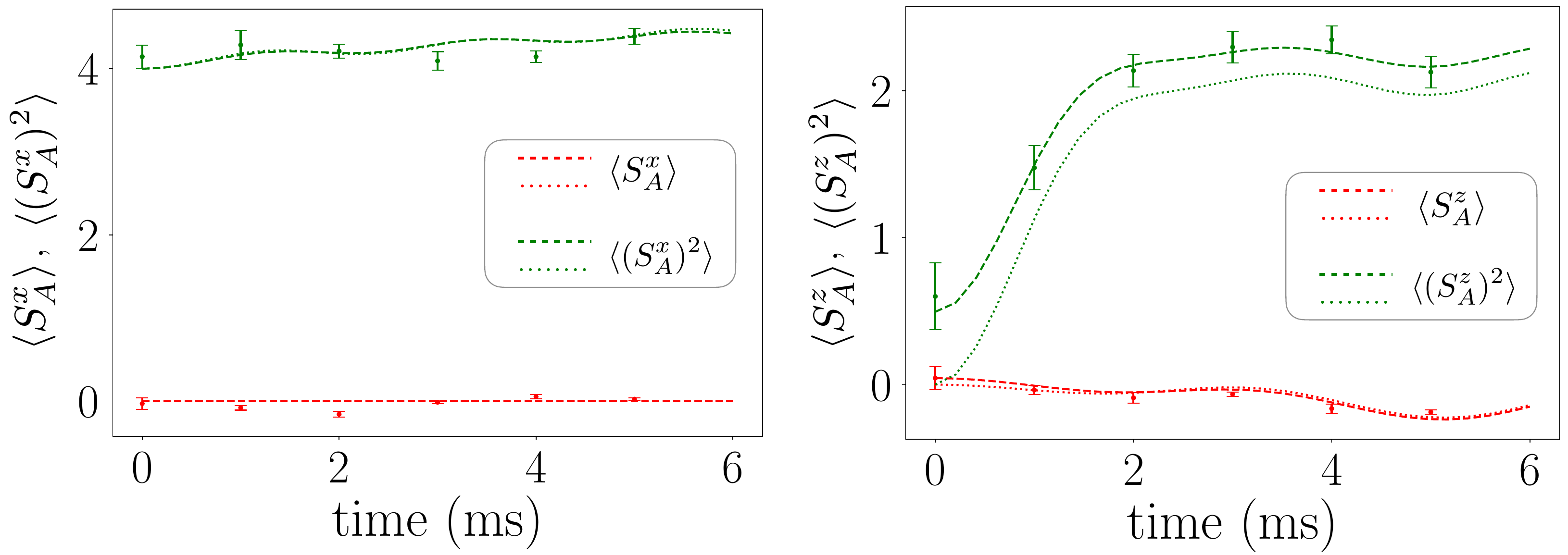}
\caption{In a quenched N\'eel state, measurements of expectation of the longitudinal magnetization and its squared  (left panel), and that of transverse magnetization and its squared  (right panel) are shown. The dotted curve show ideal theory whereas the dashes include initial state preparation and measurement errors. The system comprises  of $N=10$ qubits with subsystem $A$ sized $N_A=4$.}
\label{fig:sx-sx2-Neel}
\end{figure}
For a complementary understanding of the FCS, in the SM Fig.~\ref{fig:sx-sx2-Neel} we have presented the components $\langle S_A^x \rangle$, $\langle (S_A^x)^2 \rangle$, $\langle S_A^z \rangle$, and $\langle (S_A^z)^2 \rangle$ as functions of time. The values $\langle S_A^x \rangle$, $\langle S_A^z \rangle$ correspond directly to the mean of the correponding PDF and the slope of the imaginary FCS. Alongside, the  values $\langle (S_A^x)^2 \rangle$, $\langle (S_A^z)^2 \rangle$ with time denote the changes in the curvature of the real part of the FCS as functions of time.

\subsection{Details on the bit-flip error corrections} 
\begin{figure}
\includegraphics[width=0.38\textwidth]{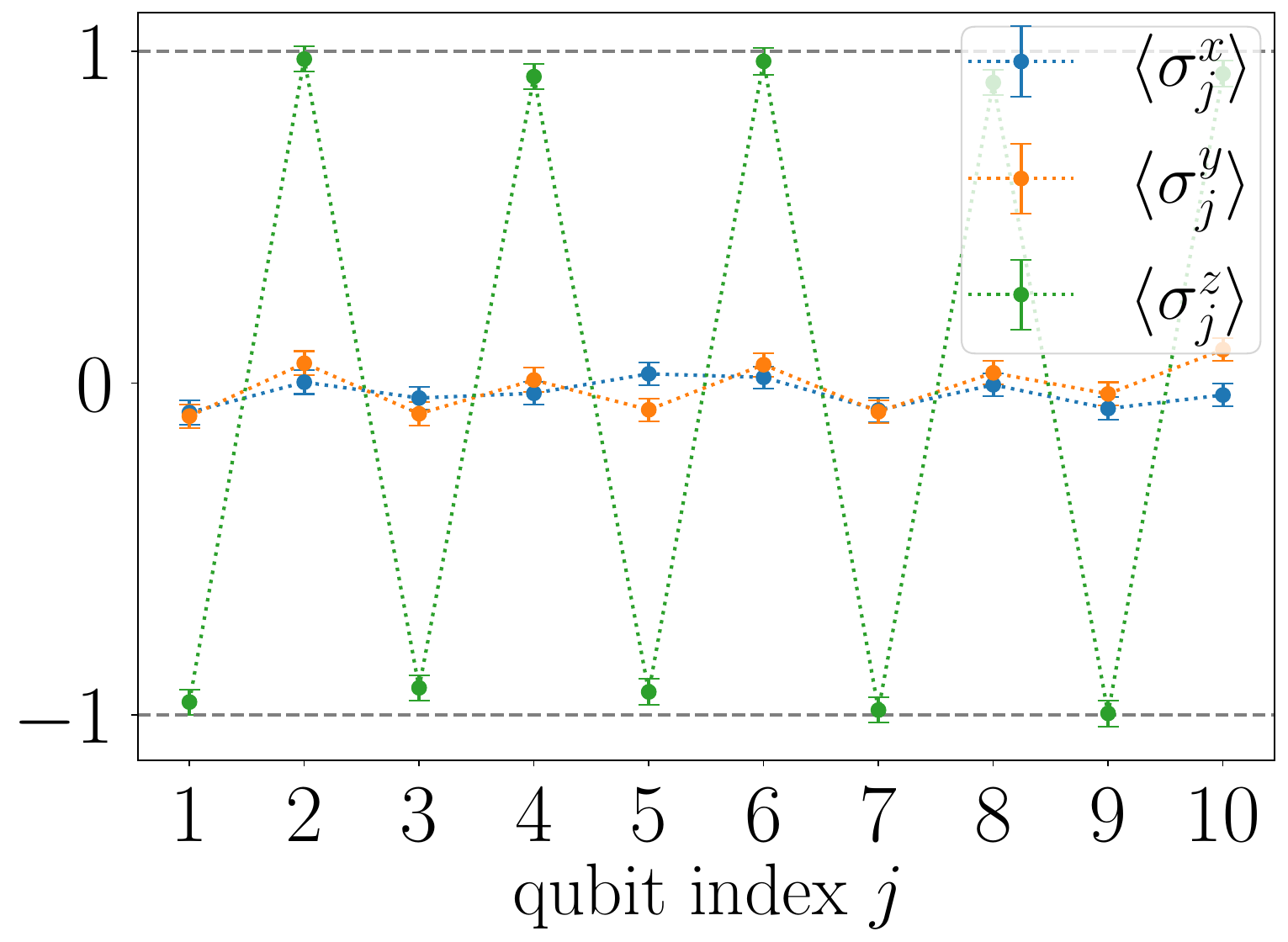}
  \caption{We show measurements of $\langle\sigma^x_j\rangle, \langle\sigma^y_j\rangle$ and $\langle\sigma^z_j\rangle$ at each site in the  $N=10$ system prepared in a N\'eel state. Classical shadows are used to obtain the expectation values. The gray dashed horizontal lines at vertical axis positions $\pm 1$ are visual guides. }
  \label{fig:Neel_xyz}
\end{figure}

In the N\'eel state, a subsystem with an even number of contiguous sites is in an eigenstate of the transverse magnetization $S_A^z$. As a result, the PDF of a measurement of $S_A^z$ is $p_{z}(q)=\delta_{q,0}$. As noted in the MT Fig.~1~(e), in the experimental data at time $t=0$, i.e., for the N\'eel state, the measured PDF is, $p_z(q=\pm 2) > 0$ and $p_z(q=0 )<1$, for $N=10$ and subsystem size $N_A=4$. In SM Fig.~\ref{fig:Neel_xyz}, we show the  measured values of $\langle\sigma^x_j\rangle, \langle\sigma^y_j\rangle$ and $\langle\sigma^z_j\rangle$ at each single site $j$. We observe a non-zero value of $\langle\sigma^{x}_j\rangle$ and $\langle\sigma^y_j\rangle$ and $|\langle\sigma^z_j\rangle|\le 1$, whereas we theoretically expect $\langle\sigma^{x}_j\rangle=\langle\sigma^y_j\rangle=0$ and  $|\langle\sigma^z_j\rangle|= 1$. This mismatch is the result of coherent as well as state preparation and measurement errors. Interestingly, in extended subsystems, the FCS and PDF for the operator $S_A^x$ (Fig.~1 MT) and the square operator $(S_A^x)^2$ (SM Fig.~\ref{fig:sx-sx2-Neel})  are not significantly affected by these experimental errors. Therefore, in modeling the experimental errors, we have only considered  decoherences due to random bit-flips in the $z$-basis, which correctly captures experimentally observed behavior for the operator $S_A^z$.

In the bit-flip model, instead of the ideal initial local qubit being  $\rho_j^{\mathrm{ideal}}=\ket{\uparrow}\bra{\uparrow}$ or $\ket{\downarrow}\bra{\downarrow}$, we assume that the experimentally realized initial state on the $j$-th qubit is
\begin{equation}
\rho_j^{\mathrm{realized}}=(1-p_j)\rho_i^{\mathrm{ideal}}+p_j \sigma^x_j\rho_j^{\mathrm{ideal}}\sigma_j^x.
\end{equation}
Here, $p_j$ is the bit-flip rate at the $j-$th qubit, which is given by $p_j=(1-\lvert\langle \sigma_j^z \rangle \rvert)/2$. We can estimate this rate in our experiment obtaining $\langle\sigma^z_j\rangle$ from the classical shadows of the reduced density matrix. The corresponding bit-flip rates $p_j$ are listed in Table~\ref{tab:rates}. Utilizing these learned bit-flip  rates, in Fig.~1~(e) in the MT, the theoretical curves for the realized state $\ket{\Psi_0}\equiv \otimes_j\rho_j^{\rm{realized}}$ are drawn in dashes.

As a side remark we note that in the ideal N\'eel state,  for even sized subsystems, the FCS is $\chi_z(\alpha)=1$. On the other hand, the FCS in the realized initial state is modified as $\chi_z(\alpha)=\prod_{j}(\cos\alpha+i(1-2p_j)\sin\alpha) \prod_{j'}(\cos\alpha-i(1-2p_{j'})\sin\alpha)$, where $j (j')$ runs over even (odd) indexed lattice sites. 

\begin{table}[h!]
\centering
\begin{tabular}{|p{0.31in}p{0.31in}| p{0.31in}p{0.31in}|}
\hline
$p_1$ &0.019 & $p_6$ &0.015 \\ \hline
$p_2$ &0.012 & $p_7$ &0.007\\ \hline
$p_3$ &0.041 & $p_8$ &0.047 \\ \hline
$p_4$ &0.038 & $p_9$ &0.002 \\ \hline
$p_5$ &0.034 & $p_{10}$ &0.034 \\ \hline 
  \end{tabular}
  \caption{Learned bit-flip rates for the N\'eel state realized in the experiment. The index $j$ in the value $p_j$ denotes the qubit site index.}%
  \label{tab:rates}
\end{table}

\subsection{FCS and PDF as functions of time}
\label{sec:timedependent_Neel}
For completeness of the results, in this section we show the measurements of the FCS and PDF as functions of time. These are the complementary figures to the ones presented in the main text. 
\subsubsection{PDF vs. time for N\'eel state}

\begin{figure}[h!]
\centering
\includegraphics[width=\linewidth]{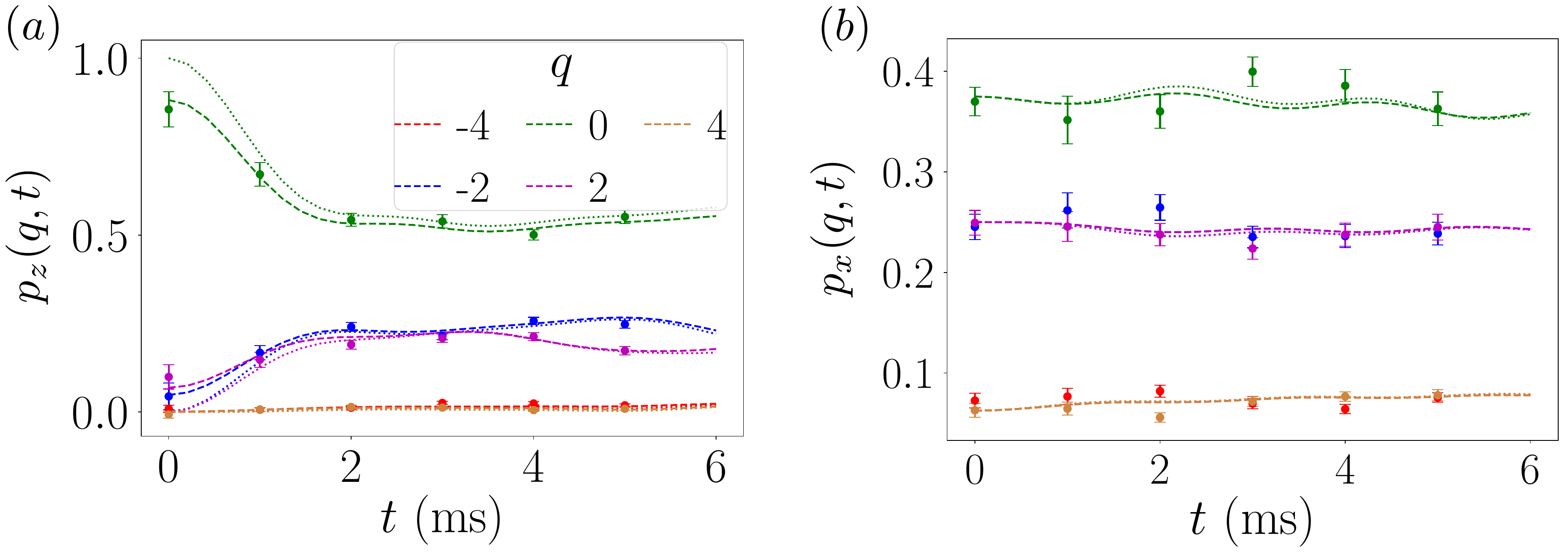}
\caption{In a system of $N=10$ qubits and $N_A=4$ sized subsystem, the probability distribution function of the operator $S_A^z$ (panel (a)) and that of the operator $S_A^x$ (panel (b)) as a function of time are shown. The measured data (dots and error bars) closely follow the theory prediction (as in dashed curves).}
\label{fig:pdf_time_neel_sx}
\end{figure}
We have studied the PDF in a quench dynamics from a N\'eel state in the main text. We present in the SM Fig.~\ref{fig:pdf_time_neel_sx} the time evolution of the PDFs $p_x(q,t)$ and $p_x(q,t)$. We notice, as also observed in Fig.~1 in the MT, that the PDF of the $x$ magnetization changes very little with time satisfying $p(q)=p(-q)$ at all times. On the other hand, the PDF of the subsystem magnetization in the $z$ direction fluctuates significantly with time. 
\subsubsection{FCS vs. time for N\'eel state}

\begin{figure}[h!]
\centering
\includegraphics[width=\linewidth]{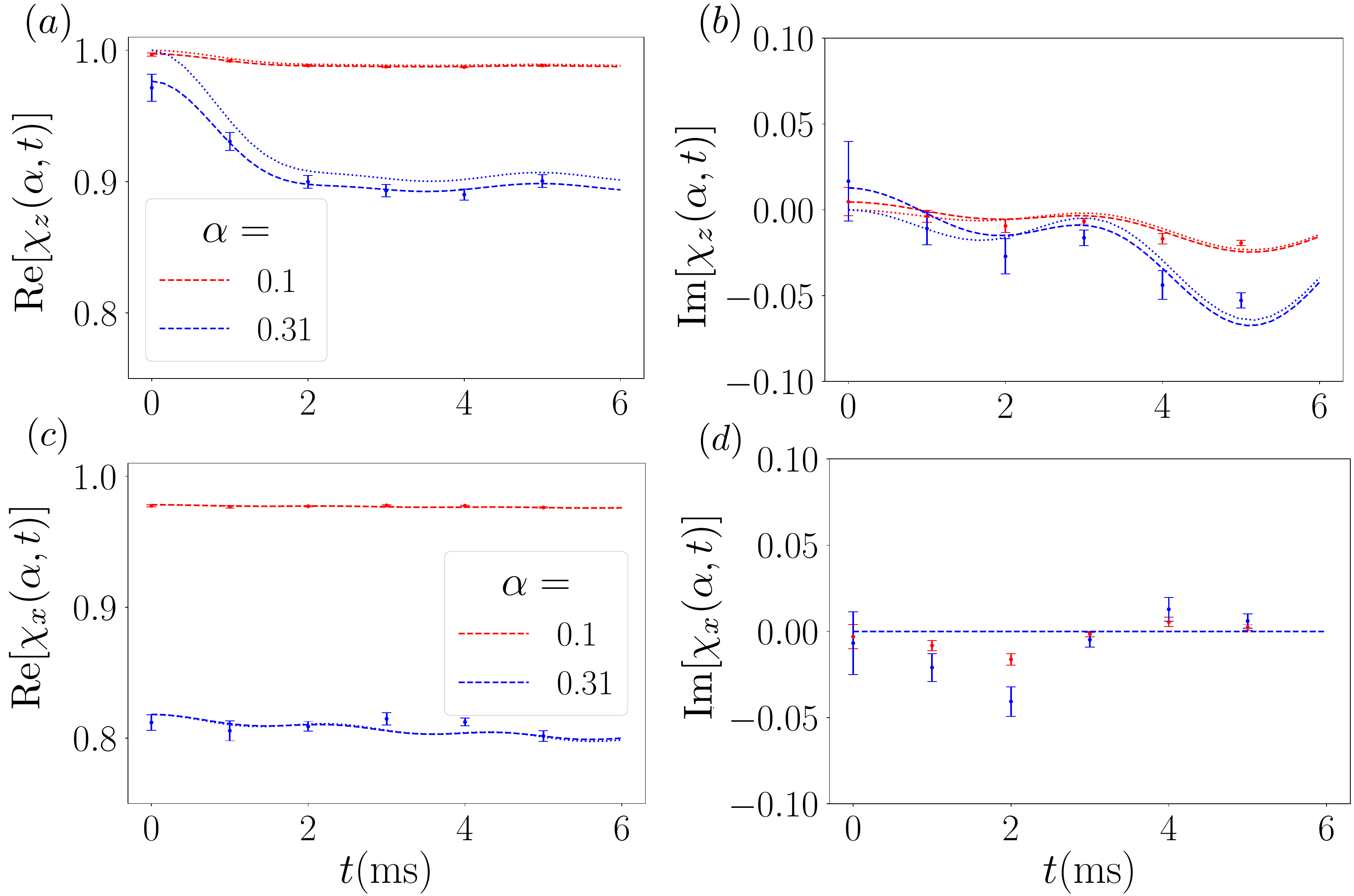}
\caption{In a system of $N=10$ qubits and $N_A=4$ sized subsystem, the full counting statistics for operators $S_A^z$ (top row) and $S_A^x$ (bottom row) are shown as a function of quenching time. The results are shown for two choices of the spectral parameter $\alpha$. The measured data (dots and error bars) closely follow the theory prediction (as in dashed curves) drawn for the error corrected initial N\'eel state.}
\label{fig:fcs_time_neel_sx_sz}
\end{figure}
The time evolution of the magnetization FCS for a N\'eel state after a quench with the XY Hamiltonian is presented in the SM Fig.~\ref{fig:fcs_time_neel_sx_sz}. In panels~(a)~and~(b), the real and imaginary parts of the FCS $\chi_z(\alpha)$ of the magnetization operator $S_A^z$ are shown, comparing the measured data and the theoretical curves. As in the main text, we note that the experimental data slightly deviate from the ideal theory (the dotted curves) during the time evolution. However, once we include the state preparation and measurement errors in the theory calculations, the curves (dashes) agree with the experimental data. In panels~(c)~and ~(d),  we measure the real and imaginary parts of the FCS $\chi_x(\alpha)$ for the magnetization $S_A^x$. The very small deviations from the ideal theory are within the allowed error due to finite measurement budget. The near zero value of the Im$[\chi_x]$ at all times is consistent with the symmetry of the PDF $p_x(q)$, as already observed in the SM Fig.~\ref{fig:pdf_time_neel_sx}(b).

\subsubsection{More on the FCS for tilted ferromagnetic state}
\label{sec:sx_ferro_more}

\begin{figure}[h!]
\centering
\includegraphics[width=\linewidth]{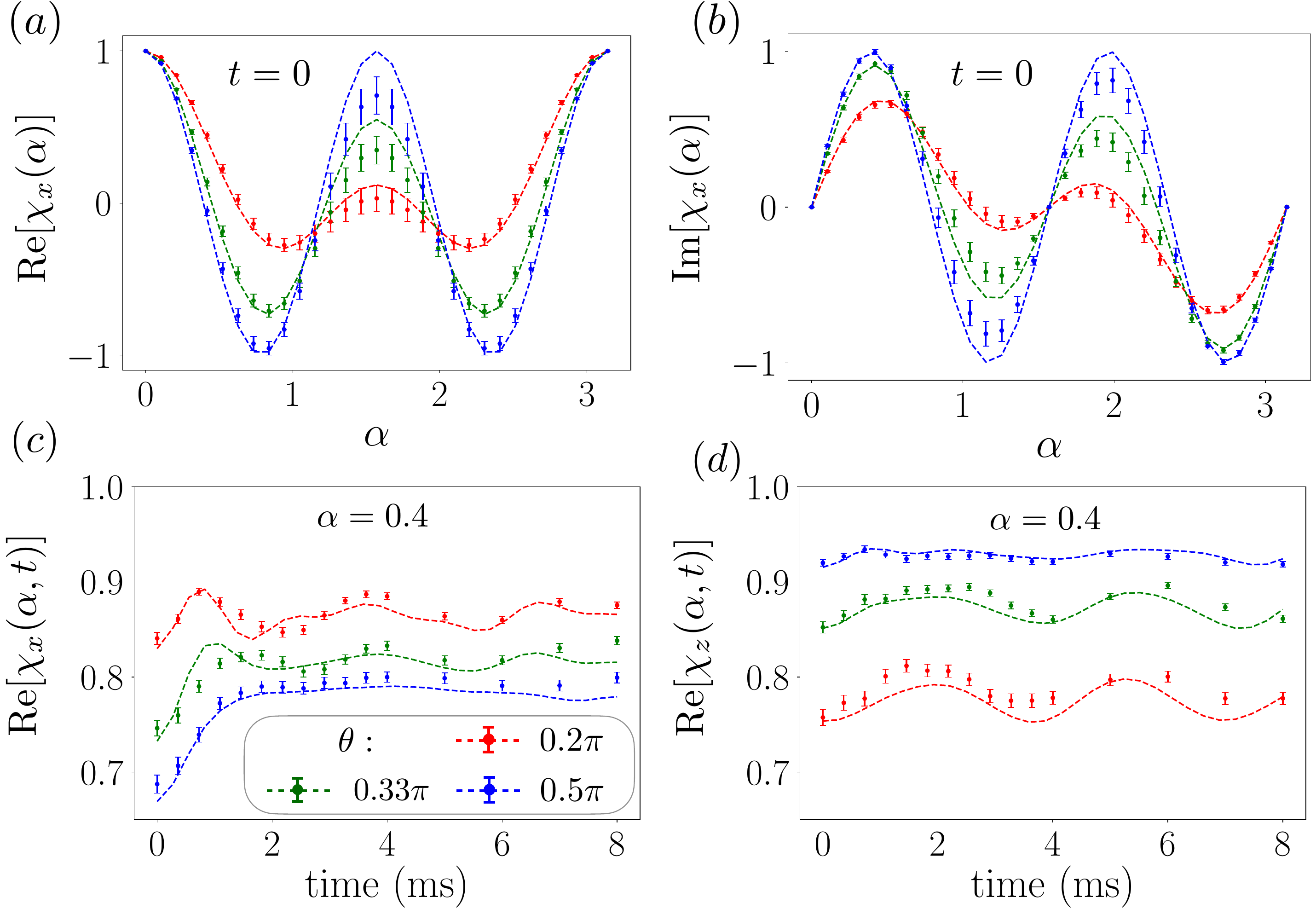}
\caption{At the initial time $t=0$, for the operator $S_A^x$ and  three tilting angles, $\theta=0.2\pi$, $0.33\pi$ and $0.5\pi$, we present the FCS vs. $\alpha$ in the top row. In the bottom row, the measurement of the real part of the FCS for the operators $S_A^x$ and $S_A^z$ as a function of time is presented for the three angles.}
\label{fig:fcs_time_alpha_Sx_sz}
\end{figure}

In SM Figs.~\ref{fig:fcs_time_alpha_Sx_sz}(a) and (b), we present the FCS  at the initial time $t=0$ for the operator $S_A^x$ and three tilting angles, $\theta=0.2\pi$, $0.33\pi$ and $0.5\pi$. The set up is the same as in Figs.~3~(a-b) in the MT, where we carry out the same analysis but for the operator $S_A ^z$. Observe that, as in that case, $\chi_x(\alpha)$ has a peak at $\alpha=\pi/2$, but now it shows the opposite behavior: it is bigger of $\theta=0.5\pi$ than for $\theta=0.3\pi$ and $0.2\pi$. As discussed in section~\ref{sec:fcsandpdf}, the reason is that $\chi_x(\pi/2)$ corresponds to the expectation value of the spin parity operator and, for a tilted ferromagnet, coincides with the component of the spins in the $x$-axis,  $\chi_x(\pi/2)=i^{N_A}(\sin\theta)^{N_A}$.

The time evolution of the real part of the FCS for the operators $S_A^x$ and $S_A^z$ during the quench dynamics is presented in the SM Figs.~\ref{fig:fcs_time_alpha_Sx_sz}~(c) and (d), respectively, complementing Figs.~3~(c) and (d) in the MT, where we plotted the corresponding imaginary parts. In the $z$-axis, as also happens with the imaginary part, it remains almost constant in time for any $\theta$. This is consistent with Fig.~2~(c) of the MT, where we found out that the corresponding PDF for $\theta=0.5\pi$ only oscillates after the quench. On the other hand, for the longitudinal magnetization, the real part of the FCS initially grows after the quench and eventually saturates around $t=2$ to a stationary value, different for each angle $\theta$ analyzed. This contrasts with the imaginary part, cf. panel~(d) of Fig.~3 in MT, which converges to zero for any $\theta$.

\subsection{Experimental parameters and error analysis} 
\label{sec:errorbars}
The experiment in Ref.~\cite{Brydges2019} begins with a N\'eel state in a system of $N=10$ ions and then a quantum quench with the XY Hamiltonian is performed. In the present work, we have used this experimental data without any change. The experiment was performed with $N_u=500$ random rotations and $N_M=150$ shots per random rotation. The experiment beginning with the tilted ferromagnetic states in Ref.~\cite{joshi-24} is done in a system of $N=12$ ions. In this case, the experiment was performed with $N_u=500$ random rotations and $N_M=30$ shots per random rotation. The experiment also employs the XY Hamiltonian for the quench and we use the data provided without any change. 

In the main text, the error bars of the PDFs are the standard error of the mean over $N_u$ random unitaries. For the FCS, the error bars show the propagation of errors due to the form of the FCS as a function of the parameter $\alpha$, see below.

\subsubsection{Error propagation in measurements of full counting statistics}
\label{sec:error_propagation}
Considering the example of the transverse magnetization operator $O_A=S_A^z=\sum_{j=1}^{N_A}\sigma_j^z$, its full counting statistics is
\begin{equation}
\chi_z(\alpha)=\tr\left(\rho_A \prod_j e^{i\alpha  \sigma_j^z}\right).
\end{equation}
Expanding this expression, we get 
\begin{eqnarray}
\chi_z(\alpha)&=&\tr[\rho_A \prod_j (\cos\alpha+i   \sigma_j^z \sin\alpha)]\nonumber\\
&=& \tr[\rho_A  ((\cos\alpha)^{N_A}+i (\cos\alpha)^{N_A-1} \sin\alpha \sum_{j=1}^{N_A}\sigma_j^z\nonumber\\
&&- (\cos\alpha )^{N_A-2}(\sin\alpha)^2 \sum_{\substack{j,j'=1\\i<j}}^{N_A}\sigma_j^z\sigma_{j'}^z\nonumber\\
&&+\dots+(i)^{N_A}\sigma_1^z\sigma_2^z\cdots \sigma_{N_A}^z (\sin\alpha)^{N_A})]~.
\end{eqnarray}
\begin{figure}[t!]
\centering
\includegraphics[width=\linewidth]{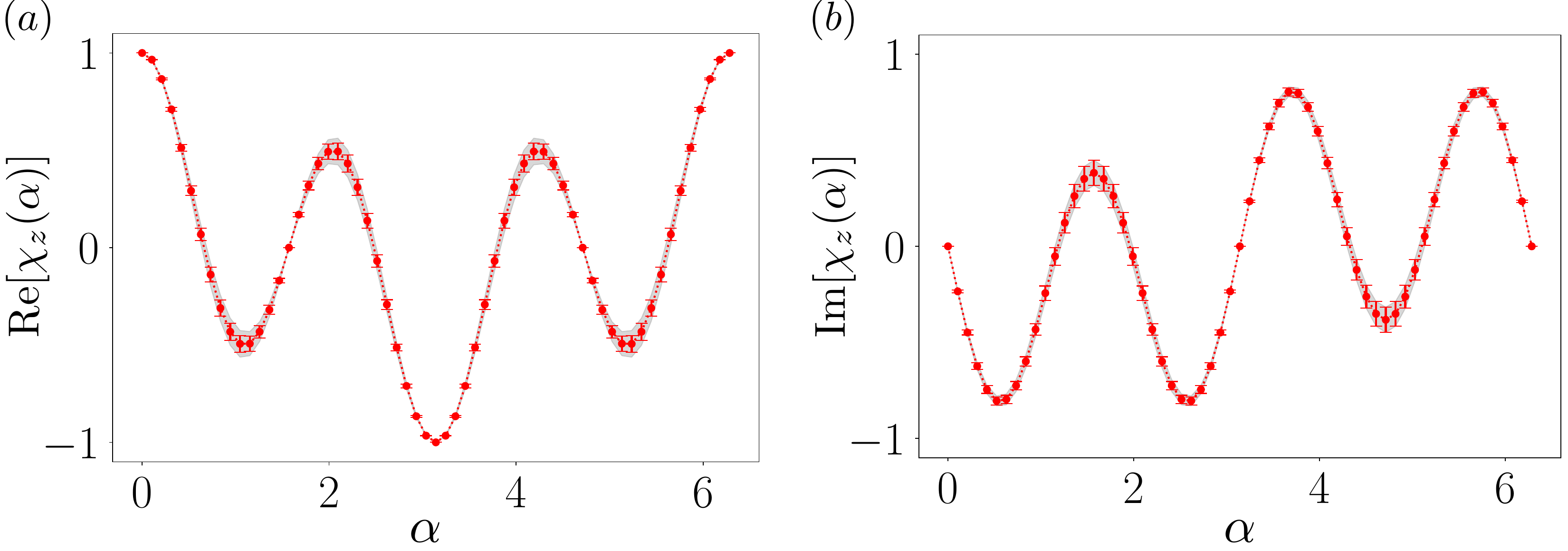}
    \caption{Real and imaginary parts of the transverse magnetization FCS for a system with $N=12$, $N_A=3$, prepared in a tilted ferromagnetic state with angle $\theta=0.2\pi$. Red dots and error bars show direct estimation as presented in the main text. Shaded region shows the experimental measurements using the expanded expression as in Eqs.~\eqref{eq:re}-\eqref{eq:error-im}. }
    \label{fig:N12Na3}
\end{figure}
The errors are correctly estimated using the propagation of uncertainties. Explicitly, taking as example a subsystem of size $N_A=3$, we have
\begin{eqnarray}
\mathrm{Re}[\chi_z(\alpha)]&=& - \cos(\alpha)\sin^2(\alpha) (\langle\sigma_1^z\sigma_2^z\rangle+\langle\sigma_2^z\sigma_3^z\rangle+\langle\sigma_1^z\sigma_3^z\rangle)\nonumber\\
&&+\cos^3(\alpha)~,
\label{eq:re}
\end{eqnarray}
and
\begin{eqnarray}
\mathrm{Im}[\chi_z(\alpha)]&=&\cos^2(\alpha) \sin(\alpha) (\langle\sigma_1^z\rangle+\langle\sigma_2^z\rangle+\langle\sigma_3^z\rangle)\nonumber\\
&& -\sin^3(\alpha) \langle\sigma_1^z\sigma_2^z\sigma_3^z\rangle~.
\label{eq:im} 
\end{eqnarray}
The corresponding errors are 
\begin{eqnarray}
   &&\eta(\mathrm{Re}[\chi(\alpha)]) \nonumber \\
&&=\cos(\alpha)\sin^2(\alpha)\left(\eta(\sigma_1^z\sigma_2^z)^2+\eta(\sigma_2^z\sigma_3^z)^2+\eta(\sigma_1^z\sigma_3^z)^2\right)^{1/2}\nonumber\\
\label{eq:error-re} 
\end{eqnarray}
and 
\begin{eqnarray}
&&\eta(\mathrm{Im}[\chi(\alpha)])= \left(\sin^3\alpha~ \eta(\sigma_1^z\sigma_2^z\sigma_3^z)^2+\right.
\nonumber\\ 
&&\left.\cos^2\alpha \sin\alpha (\eta(\sigma_1^z)^2+\eta(\sigma_2^z)^2+\eta(\sigma_3^z)^2)\right)^{1/2}~,
\label{eq:error-im}
\end{eqnarray}
where $\eta(O)$ denotes the error in the estimation of the observable $O$. Note that, the above expression is strictly valid for a completely independent measurement of each term in the expansion. In the SM Fig.~\ref{fig:N12Na3}, we take an example of a system of $N=12$ spins prepared in a tilted ferromagnetic state $|\Psi_0(\theta)\rangle = e^{i\theta \sum \sigma_j^y/2}|\downarrow\downarrow\dots\rangle$, with $\theta=0.2\pi$, and use the data from the experiment to measure the FCS. The red dots and error bars are obtained using the direct estimation as presented in the main text, whereas the shaded region is drawn using the formulas presented in this section, using the same data set to obtain quantities like $\langle \sigma^z_1\rangle$, $\langle \sigma^z_1\sigma^z_2\rangle$, etc. We notice that our formulas correctly capture the qualitative behavior of the error bars propagation as a function of $\alpha$. Note also that, since the subsystem size is odd $N_A=3$, the period of the FCS is $\alpha=2\pi$ instead of $\alpha=\pi$, as seen in the main text for $N_A=4$.

\subsection{Average probabilities}
\label{sec:average_prob}

\begin{figure}[t]
\centering
\includegraphics[width=0.7\linewidth]{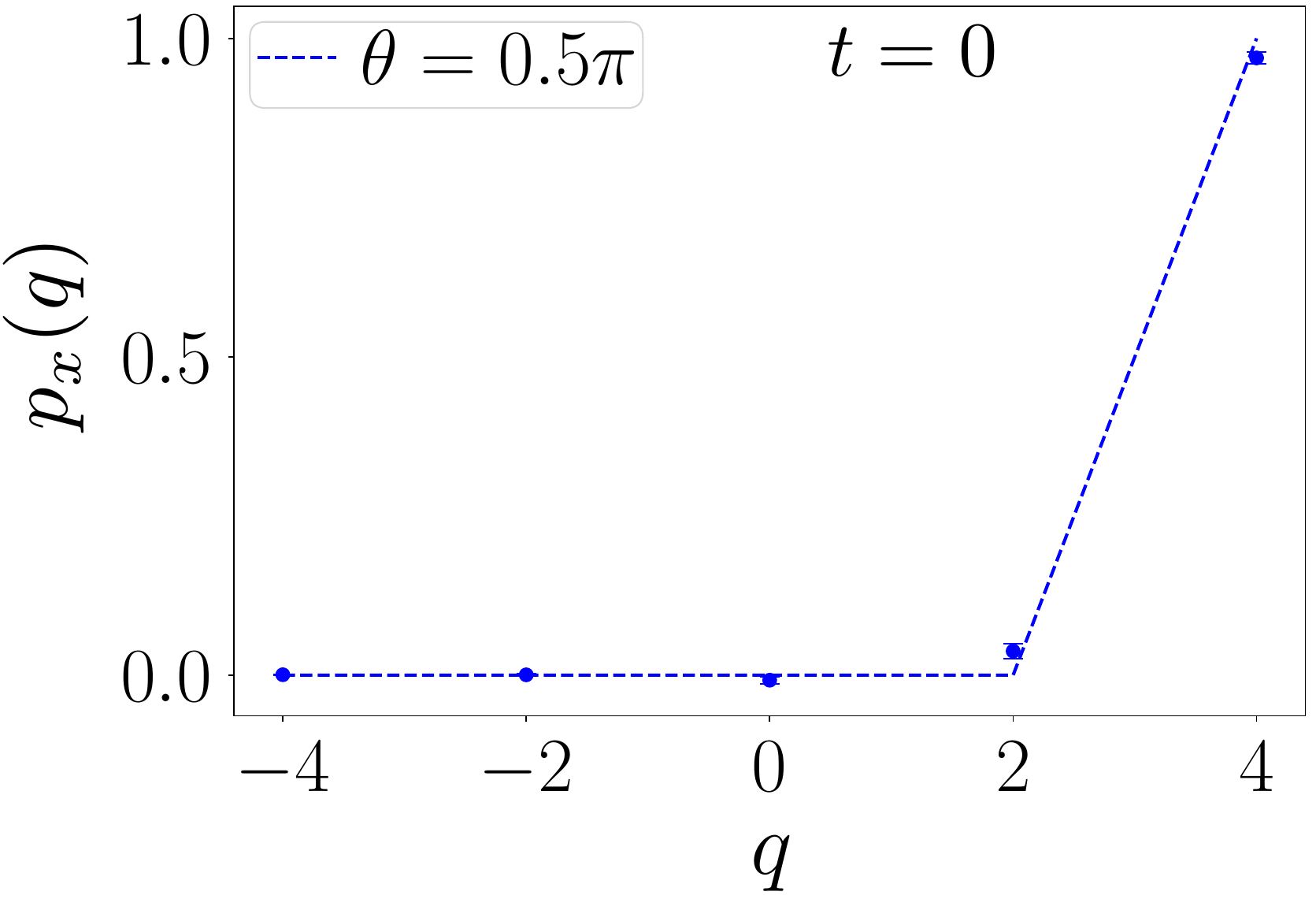}
\caption{Average probabilities in a measurement of $S_A^x$ over all bulk subsystems of size $N_A=4$ for a tilted ferromagnetic state at angle $\theta=0.5\pi$. }
\label{fig:probs_t0_averaged}
\end{figure}

To reduce statistical error, we can consider taking the average over all subsystems of the same size in the bulk. As an example, in SM Fig.~\ref{fig:probs_t0_averaged}, we show the measurement of the PDF $p_x(q)$ at $t=0$ for the tilted ferromagnetic state $\vert \Psi(\theta=0.5\pi) \rangle$ in a system of 12 qubits with subsystems of size $N_A=4$. We notice that, compared to the PDF for a single interval presented in the main text, cf. MT Fig.~2~(b), the statistical errors are significantly reduced. However, there is still some measurement error. This error is difficult to model since the experimental measurements are tailored to specific measurement basis, with focus to reduce local $z-$rotations during measurements. Nevertheless, the measurement sequence is performed uniformly to realize the Haar random unitaries.

\bibliography{Bib_QMPE, Bib_FCS}

\begin{thebibliography}{85}%
\makeatletter
\providecommand \@ifxundefined [1]{%
 \@ifx{#1\undefined}
}%
\providecommand \@ifnum [1]{%
 \ifnum #1\expandafter \@firstoftwo
 \else \expandafter \@secondoftwo
 \fi
}%
\providecommand \@ifx [1]{%
 \ifx #1\expandafter \@firstoftwo
 \else \expandafter \@secondoftwo
 \fi
}%
\providecommand \natexlab [1]{#1}%
\providecommand \enquote  [1]{``#1''}%
\providecommand \bibnamefont  [1]{#1}%
\providecommand \bibfnamefont [1]{#1}%
\providecommand \citenamefont [1]{#1}%
\providecommand \href@noop [0]{\@secondoftwo}%
\providecommand \href [0]{\begingroup \@sanitize@url \@href}%
\providecommand \@href[1]{\@@startlink{#1}\@@href}%
\providecommand \@@href[1]{\endgroup#1\@@endlink}%
\providecommand \@sanitize@url [0]{\catcode `\\12\catcode `\$12\catcode
  `\&12\catcode `\#12\catcode `\^12\catcode `\_12\catcode `\%12\relax}%
\providecommand \@@startlink[1]{}%
\providecommand \@@endlink[0]{}%
\providecommand \url  [0]{\begingroup\@sanitize@url \@url }%
\providecommand \@url [1]{\endgroup\@href {#1}{\urlprefix }}%
\providecommand \urlprefix  [0]{URL }%
\providecommand \Eprint [0]{\href }%
\providecommand \doibase [0]{http://dx.doi.org/}%
\providecommand \selectlanguage [0]{\@gobble}%
\providecommand \bibinfo  [0]{\@secondoftwo}%
\providecommand \bibfield  [0]{\@secondoftwo}%
\providecommand \translation [1]{[#1]}%
\providecommand \BibitemOpen [0]{}%
\providecommand \bibitemStop [0]{}%
\providecommand \bibitemNoStop [0]{.\EOS\space}%
\providecommand \EOS [0]{\spacefactor3000\relax}%
\providecommand \BibitemShut  [1]{\csname bibitem#1\endcsname}%
\let\auto@bib@innerbib\@empty
\bibitem [{\citenamefont {Blanter}\ and\ \citenamefont
  {Buttiker}(2000)}]{blanter-00}%
  \BibitemOpen
  \bibfield  {author} {\bibinfo {author} {\bibfnamefont {Y.~M.}\ \bibnamefont
  {Blanter}}\ and\ \bibinfo {author} {\bibfnamefont {M.}~\bibnamefont
  {Buttiker}},\ }\href {https://doi.org/10.1016/S0370-1573%2899%2900123-4}
  {\bibfield  {journal} {\bibinfo  {journal} {Phys. Rep.}\ }\textbf {\bibinfo
  {volume} {336}},\ \bibinfo {pages} {1} (\bibinfo {year} {2000})}\BibitemShut
  {NoStop}%
\bibitem [{\citenamefont {Gardiner}\ and\ \citenamefont
  {Zoller}(2004)}]{gardiner}%
  \BibitemOpen
  \bibfield  {author} {\bibinfo {author} {\bibfnamefont {C.}~\bibnamefont
  {Gardiner}}\ and\ \bibinfo {author} {\bibfnamefont {P.}~\bibnamefont
  {Zoller}},\ }\enquote {\bibinfo {title} {Quantum noise: A handbook of
  markovian and non-markovian quantum stochastic methods with applications to
  quantum optics},}\ \ (\bibinfo  {publisher} {Springer-Verlag Berlin
  Heidelberg},\ \bibinfo {year} {2004})\BibitemShut {NoStop}%
\bibitem [{\citenamefont {Nazarov}(2012)}]{nazarov}%
  \BibitemOpen
  \bibfield  {author} {\bibinfo {author} {\bibfnamefont {Y.~V.}\ \bibnamefont
  {Nazarov}},\ }\enquote {\bibinfo {title} {Quantum noise in mesoscopic
  physics},}\ \ (\bibinfo  {publisher} {Springer Dordrecht},\ \bibinfo {year}
  {2012})\BibitemShut {NoStop}%
\bibitem [{\citenamefont {Lamb}\ and\ \citenamefont
  {Retherford}(1947)}]{Lamb_1947}%
  \BibitemOpen
  \bibfield  {author} {\bibinfo {author} {\bibfnamefont {W.~E.}\ \bibnamefont
  {Lamb}}\ and\ \bibinfo {author} {\bibfnamefont {R.~C.}\ \bibnamefont
  {Retherford}},\ }\href {\doibase 10.1103/PhysRev.72.241} {\bibfield
  {journal} {\bibinfo  {journal} {Phys. Rev.}\ }\textbf {\bibinfo {volume}
  {72}},\ \bibinfo {pages} {241} (\bibinfo {year} {1947})}\BibitemShut
  {NoStop}%
\bibitem [{\citenamefont {Bellwied}\ \emph {et~al.}(2015)\citenamefont
  {Bellwied}, \citenamefont {Bors\'anyi}, \citenamefont {Fodor}, \citenamefont
  {Katz}, \citenamefont {P\'asztor}, \citenamefont {Ratti},\ and\ \citenamefont
  {Szab\'o}}]{QCD_Bellwied}%
  \BibitemOpen
  \bibfield  {author} {\bibinfo {author} {\bibfnamefont {R.}~\bibnamefont
  {Bellwied}}, \bibinfo {author} {\bibfnamefont {S.}~\bibnamefont
  {Bors\'anyi}}, \bibinfo {author} {\bibfnamefont {Z.}~\bibnamefont {Fodor}},
  \bibinfo {author} {\bibfnamefont {S.~D.}\ \bibnamefont {Katz}}, \bibinfo
  {author} {\bibfnamefont {A.}~\bibnamefont {P\'asztor}}, \bibinfo {author}
  {\bibfnamefont {C.}~\bibnamefont {Ratti}}, \ and\ \bibinfo {author}
  {\bibfnamefont {K.~K.}\ \bibnamefont {Szab\'o}},\ }\href {\doibase
  10.1103/PhysRevD.92.114505} {\bibfield  {journal} {\bibinfo  {journal} {Phys.
  Rev. D}\ }\textbf {\bibinfo {volume} {92}},\ \bibinfo {pages} {114505}
  (\bibinfo {year} {2015})}\BibitemShut {NoStop}%
\bibitem [{\citenamefont {Herrero-Collantes}\ and\ \citenamefont
  {Garcia-Escartin}(2017)}]{QRNG_RevModPhys}%
  \BibitemOpen
  \bibfield  {author} {\bibinfo {author} {\bibfnamefont {M.}~\bibnamefont
  {Herrero-Collantes}}\ and\ \bibinfo {author} {\bibfnamefont {J.~C.}\
  \bibnamefont {Garcia-Escartin}},\ }\href {\doibase
  10.1103/RevModPhys.89.015004} {\bibfield  {journal} {\bibinfo  {journal}
  {Rev. Mod. Phys.}\ }\textbf {\bibinfo {volume} {89}},\ \bibinfo {pages}
  {015004} (\bibinfo {year} {2017})}\BibitemShut {NoStop}%
\bibitem [{\citenamefont {{Lamoreaux}}(2005)}]{Casimir}%
  \BibitemOpen
  \bibfield  {author} {\bibinfo {author} {\bibfnamefont {S.~K.}\ \bibnamefont
  {{Lamoreaux}}},\ }\href {\doibase 10.1088/0034-4885/68/1/R04} {\bibfield
  {journal} {\bibinfo  {journal} {Reports on Progress in Physics}\ }\textbf
  {\bibinfo {volume} {68}},\ \bibinfo {pages} {201} (\bibinfo {year}
  {2005})}\BibitemShut {NoStop}%
\bibitem [{\citenamefont {Hofferberth}\ \emph {et~al.}(2008)\citenamefont
  {Hofferberth}, \citenamefont {I.~Lesanovsky}, \citenamefont {Imambekov},
  \citenamefont {Gritsev}, \citenamefont {Demler},\ and\ \citenamefont
  {Schmiedmayer}}]{hofferberth-08}%
  \BibitemOpen
  \bibfield  {author} {\bibinfo {author} {\bibfnamefont {S.}~\bibnamefont
  {Hofferberth}}, \bibinfo {author} {\bibfnamefont {T.~S.}\ \bibnamefont
  {I.~Lesanovsky}}, \bibinfo {author} {\bibfnamefont {A.}~\bibnamefont
  {Imambekov}}, \bibinfo {author} {\bibfnamefont {V.}~\bibnamefont {Gritsev}},
  \bibinfo {author} {\bibfnamefont {E.}~\bibnamefont {Demler}}, \ and\ \bibinfo
  {author} {\bibfnamefont {J.}~\bibnamefont {Schmiedmayer}},\ }\href
  {https://doi.org/10.1038/nphys941} {\bibfield  {journal} {\bibinfo  {journal}
  {Nature Phys.}\ }\textbf {\bibinfo {volume} {4}},\ \bibinfo {pages} {489}
  (\bibinfo {year} {2008})}\BibitemShut {NoStop}%
\bibitem [{\citenamefont {Armijo}\ \emph {et~al.}(2010)\citenamefont {Armijo},
  \citenamefont {Jacqmin}, \citenamefont {Kheruntsyan},\ and\ \citenamefont
  {Bouchoule}}]{armijo-10}%
  \BibitemOpen
  \bibfield  {author} {\bibinfo {author} {\bibfnamefont {J.}~\bibnamefont
  {Armijo}}, \bibinfo {author} {\bibfnamefont {T.}~\bibnamefont {Jacqmin}},
  \bibinfo {author} {\bibfnamefont {K.~V.}\ \bibnamefont {Kheruntsyan}}, \ and\
  \bibinfo {author} {\bibfnamefont {I.}~\bibnamefont {Bouchoule}},\ }\href
  {https://doi.org/10.1103/PhysRevLett.105.230402} {\bibfield  {journal}
  {\bibinfo  {journal} {Phys. Rev. Lett.}\ }\textbf {\bibinfo {volume} {105}},\
  \bibinfo {pages} {230402} (\bibinfo {year} {2010})}\BibitemShut {NoStop}%
\bibitem [{\citenamefont {Jacqmin}\ \emph {et~al.}(2011)\citenamefont
  {Jacqmin}, \citenamefont {Armijo}, \citenamefont {Berrada}, \citenamefont
  {Kheruntsyan},\ and\ \citenamefont {Bouchoule}}]{jacqmin-11}%
  \BibitemOpen
  \bibfield  {author} {\bibinfo {author} {\bibfnamefont {T.}~\bibnamefont
  {Jacqmin}}, \bibinfo {author} {\bibfnamefont {J.}~\bibnamefont {Armijo}},
  \bibinfo {author} {\bibfnamefont {T.}~\bibnamefont {Berrada}}, \bibinfo
  {author} {\bibfnamefont {K.~V.}\ \bibnamefont {Kheruntsyan}}, \ and\ \bibinfo
  {author} {\bibfnamefont {I.}~\bibnamefont {Bouchoule}},\ }\href
  {https://doi.org/10.1103/PhysRevLett.106.230405} {\bibfield  {journal}
  {\bibinfo  {journal} {Phys. Rev. Lett.}\ }\textbf {\bibinfo {volume} {106}},\
  \bibinfo {pages} {230405} (\bibinfo {year} {2011})}\BibitemShut {NoStop}%
\bibitem [{\citenamefont {Gring}\ \emph {et~al.}(2012)\citenamefont {Gring},
  \citenamefont {Kuhnert}, \citenamefont {Langen}, \citenamefont {Kitagawa},
  \citenamefont {Rauer}, \citenamefont {Schreitl}, \citenamefont {Mazets},
  \citenamefont {Smith}, \citenamefont {Demler},\ and\ \citenamefont
  {Schmiedmayer}}]{gring-12}%
  \BibitemOpen
  \bibfield  {author} {\bibinfo {author} {\bibfnamefont {M.}~\bibnamefont
  {Gring}}, \bibinfo {author} {\bibfnamefont {M.}~\bibnamefont {Kuhnert}},
  \bibinfo {author} {\bibfnamefont {T.}~\bibnamefont {Langen}}, \bibinfo
  {author} {\bibfnamefont {T.}~\bibnamefont {Kitagawa}}, \bibinfo {author}
  {\bibfnamefont {B.}~\bibnamefont {Rauer}}, \bibinfo {author} {\bibfnamefont
  {M.}~\bibnamefont {Schreitl}}, \bibinfo {author} {\bibfnamefont
  {I.}~\bibnamefont {Mazets}}, \bibinfo {author} {\bibfnamefont {D.~A.}\
  \bibnamefont {Smith}}, \bibinfo {author} {\bibfnamefont {E.}~\bibnamefont
  {Demler}}, \ and\ \bibinfo {author} {\bibfnamefont {J.}~\bibnamefont
  {Schmiedmayer}},\ }\href {https://doi.org/10.1126/science.1224953} {\bibfield
   {journal} {\bibinfo  {journal} {Science}\ }\textbf {\bibinfo {volume}
  {337}},\ \bibinfo {pages} {1318} (\bibinfo {year} {2012})}\BibitemShut
  {NoStop}%
\bibitem [{\citenamefont {Bohnet}\ \emph {et~al.}(2016)\citenamefont {Bohnet},
  \citenamefont {Sawyer}, \citenamefont {Britton}, \citenamefont {Wall},
  \citenamefont {Rey}, \citenamefont {Foss-Feig},\ and\ \citenamefont
  {Bollinger}}]{bohnet-16}%
  \BibitemOpen
  \bibfield  {author} {\bibinfo {author} {\bibfnamefont {J.~G.}\ \bibnamefont
  {Bohnet}}, \bibinfo {author} {\bibfnamefont {B.~C.}\ \bibnamefont {Sawyer}},
  \bibinfo {author} {\bibfnamefont {J.~W.}\ \bibnamefont {Britton}}, \bibinfo
  {author} {\bibfnamefont {M.~L.}\ \bibnamefont {Wall}}, \bibinfo {author}
  {\bibfnamefont {A.~M.}\ \bibnamefont {Rey}}, \bibinfo {author} {\bibfnamefont
  {M.}~\bibnamefont {Foss-Feig}}, \ and\ \bibinfo {author} {\bibfnamefont
  {J.~J.}\ \bibnamefont {Bollinger}},\ }\href
  {https://doi.org/10.1126/science.aad9958} {\bibfield  {journal} {\bibinfo
  {journal} {Science}\ }\textbf {\bibinfo {volume} {352}},\ \bibinfo {pages}
  {1297} (\bibinfo {year} {2016})}\BibitemShut {NoStop}%
\bibitem [{\citenamefont {Rispoli}\ \emph {et~al.}(2019)\citenamefont
  {Rispoli}, \citenamefont {Lukin}, \citenamefont {Schittko}, \citenamefont
  {Kim}, \citenamefont {Tai}, \citenamefont {Leonard},\ and\ \citenamefont
  {Greiner}}]{rispoli-19}%
  \BibitemOpen
  \bibfield  {author} {\bibinfo {author} {\bibfnamefont {M.}~\bibnamefont
  {Rispoli}}, \bibinfo {author} {\bibfnamefont {A.}~\bibnamefont {Lukin}},
  \bibinfo {author} {\bibfnamefont {R.}~\bibnamefont {Schittko}}, \bibinfo
  {author} {\bibfnamefont {S.}~\bibnamefont {Kim}}, \bibinfo {author}
  {\bibfnamefont {M.~E.}\ \bibnamefont {Tai}}, \bibinfo {author} {\bibfnamefont
  {J.}~\bibnamefont {Leonard}}, \ and\ \bibinfo {author} {\bibfnamefont
  {M.}~\bibnamefont {Greiner}},\ }\href
  {https://doi.org/10.1038/s41586-019-1527-2} {\bibfield  {journal} {\bibinfo
  {journal} {Nature}\ }\textbf {\bibinfo {volume} {573}},\ \bibinfo {pages}
  {385} (\bibinfo {year} {2019})}\BibitemShut {NoStop}%
\bibitem [{\citenamefont {Cherng}\ and\ \citenamefont
  {Demler}(2007)}]{cherng-07}%
  \BibitemOpen
  \bibfield  {author} {\bibinfo {author} {\bibfnamefont {R.~W.}\ \bibnamefont
  {Cherng}}\ and\ \bibinfo {author} {\bibfnamefont {E.}~\bibnamefont
  {Demler}},\ }\href
  {https://iopscience.iop.org/article/10.1088/1367-2630/9/1/007} {\bibfield
  {journal} {\bibinfo  {journal} {New J. Phys.}\ }\textbf {\bibinfo {volume}
  {9}},\ \bibinfo {pages} {7} (\bibinfo {year} {2007})}\BibitemShut {NoStop}%
\bibitem [{\citenamefont {Bortz}\ \emph {et~al.}(2007)\citenamefont {Bortz},
  \citenamefont {Sato},\ and\ \citenamefont {Shiroishi}}]{bortz-07}%
  \BibitemOpen
  \bibfield  {author} {\bibinfo {author} {\bibfnamefont {M.}~\bibnamefont
  {Bortz}}, \bibinfo {author} {\bibfnamefont {J.}~\bibnamefont {Sato}}, \ and\
  \bibinfo {author} {\bibfnamefont {M.}~\bibnamefont {Shiroishi}},\ }\href
  {https://iopscience.iop.org/article/10.1088/1751-8113/40/16/001} {\bibfield
  {journal} {\bibinfo  {journal} {J. Phys. A: Math. Theor.}\ }\textbf {\bibinfo
  {volume} {40}},\ \bibinfo {pages} {4253} (\bibinfo {year}
  {2007})}\BibitemShut {NoStop}%
\bibitem [{\citenamefont {Abraham}\ \emph {et~al.}(2007)\citenamefont
  {Abraham}, \citenamefont {Essler},\ and\ \citenamefont
  {Maciolek}}]{abraham-07}%
  \BibitemOpen
  \bibfield  {author} {\bibinfo {author} {\bibfnamefont {D.~B.}\ \bibnamefont
  {Abraham}}, \bibinfo {author} {\bibfnamefont {F.~H.~L.}\ \bibnamefont
  {Essler}}, \ and\ \bibinfo {author} {\bibfnamefont {A.}~\bibnamefont
  {Maciolek}},\ }\href {https://doi.org/10.1103/PhysRevLett.98.170602}
  {\bibfield  {journal} {\bibinfo  {journal} {Phys. Rev. Lett.}\ }\textbf
  {\bibinfo {volume} {98}},\ \bibinfo {pages} {170602} (\bibinfo {year}
  {2007})}\BibitemShut {NoStop}%
\bibitem [{\citenamefont {Lamacraft}\ and\ \citenamefont
  {Fendley}(2008)}]{lamacraft-08}%
  \BibitemOpen
  \bibfield  {author} {\bibinfo {author} {\bibfnamefont {A.}~\bibnamefont
  {Lamacraft}}\ and\ \bibinfo {author} {\bibfnamefont {P.}~\bibnamefont
  {Fendley}},\ }\href {https://doi.org/10.1103/PhysRevLett.100.165706}
  {\bibfield  {journal} {\bibinfo  {journal} {Phys. Rev. Lett.}\ }\textbf
  {\bibinfo {volume} {100}},\ \bibinfo {pages} {165706} (\bibinfo {year}
  {2008})}\BibitemShut {NoStop}%
\bibitem [{\citenamefont {Ivanov}\ and\ \citenamefont
  {Abanov}(2013)}]{ivanov-13}%
  \BibitemOpen
  \bibfield  {author} {\bibinfo {author} {\bibfnamefont {D.~A.}\ \bibnamefont
  {Ivanov}}\ and\ \bibinfo {author} {\bibfnamefont {A.~G.}\ \bibnamefont
  {Abanov}},\ }\href {https://doi.org/10.1103/PhysRevE.87.022114} {\bibfield
  {journal} {\bibinfo  {journal} {Phys. Rev. E}\ }\textbf {\bibinfo {volume}
  {87}},\ \bibinfo {pages} {022114} (\bibinfo {year} {2013})}\BibitemShut
  {NoStop}%
\bibitem [{\citenamefont {Eisler}(2013)}]{eisler-13}%
  \BibitemOpen
  \bibfield  {author} {\bibinfo {author} {\bibfnamefont {V.}~\bibnamefont
  {Eisler}},\ }\href {https://doi.org/10.1103/PhysRevLett.111.080402}
  {\bibfield  {journal} {\bibinfo  {journal} {Phys. Rev. Lett.}\ }\textbf
  {\bibinfo {volume} {111}},\ \bibinfo {pages} {080402} (\bibinfo {year}
  {2013})}\BibitemShut {NoStop}%
\bibitem [{\citenamefont {Klich}(2014)}]{klich-14}%
  \BibitemOpen
  \bibfield  {author} {\bibinfo {author} {\bibfnamefont {I.}~\bibnamefont
  {Klich}},\ }\href {http://dx.doi.org/10.1088/1742-5468/2014/11/P11006}
  {\bibfield  {journal} {\bibinfo  {journal} {J. Stat. Mech.}\ ,\ \bibinfo
  {pages} {P11006}} (\bibinfo {year} {2014})}\BibitemShut {NoStop}%
\bibitem [{\citenamefont {Moreno-Cardoner}\ \emph {et~al.}(2016)\citenamefont
  {Moreno-Cardoner}, \citenamefont {Sherson},\ and\ \citenamefont
  {Chiara}}]{moreno-16}%
  \BibitemOpen
  \bibfield  {author} {\bibinfo {author} {\bibfnamefont {M.}~\bibnamefont
  {Moreno-Cardoner}}, \bibinfo {author} {\bibfnamefont {J.~F.}\ \bibnamefont
  {Sherson}}, \ and\ \bibinfo {author} {\bibfnamefont {G.~D.}\ \bibnamefont
  {Chiara}},\ }\href {http://dx.doi.org/10.1088/1742-5468/2014/11/P11006}
  {\bibfield  {journal} {\bibinfo  {journal} {New J. Phys.}\ }\textbf {\bibinfo
  {volume} {18}},\ \bibinfo {pages} {103015} (\bibinfo {year}
  {2016})}\BibitemShut {NoStop}%
\bibitem [{\citenamefont {Najafi}\ and\ \citenamefont
  {Rajabpour}(2017)}]{najafi-17}%
  \BibitemOpen
  \bibfield  {author} {\bibinfo {author} {\bibfnamefont {K.}~\bibnamefont
  {Najafi}}\ and\ \bibinfo {author} {\bibfnamefont {M.}~\bibnamefont
  {Rajabpour}},\ }\href {https://doi.org/10.1103/PhysRevB.96.235109} {\bibfield
   {journal} {\bibinfo  {journal} {Phys. Rev. B}\ }\textbf {\bibinfo {volume}
  {96}},\ \bibinfo {pages} {235109} (\bibinfo {year} {2017})}\BibitemShut
  {NoStop}%
\bibitem [{\citenamefont {St\'ephan}\ and\ \citenamefont
  {Pollmann}(2017)}]{stephan-17}%
  \BibitemOpen
  \bibfield  {author} {\bibinfo {author} {\bibfnamefont {J.-M.}\ \bibnamefont
  {St\'ephan}}\ and\ \bibinfo {author} {\bibfnamefont {F.}~\bibnamefont
  {Pollmann}},\ }\href {https://doi.org/10.1103/PhysRevB.95.035119} {\bibfield
  {journal} {\bibinfo  {journal} {Phys. Rev. B}\ }\textbf {\bibinfo {volume}
  {95}},\ \bibinfo {pages} {035119} (\bibinfo {year} {2017})}\BibitemShut
  {NoStop}%
\bibitem [{\citenamefont {Collura}\ \emph {et~al.}(2017)\citenamefont
  {Collura}, \citenamefont {Essler},\ and\ \citenamefont {Groha}}]{collura-17}%
  \BibitemOpen
  \bibfield  {author} {\bibinfo {author} {\bibfnamefont {M.}~\bibnamefont
  {Collura}}, \bibinfo {author} {\bibfnamefont {F.~H.~L.}\ \bibnamefont
  {Essler}}, \ and\ \bibinfo {author} {\bibfnamefont {S.}~\bibnamefont
  {Groha}},\ }\href
  {https://iopscience.iop.org/article/10.1088/1751-8121/aa87dd/pdf} {\bibfield
  {journal} {\bibinfo  {journal} {J. Phys. A}\ }\textbf {\bibinfo {volume}
  {50}},\ \bibinfo {pages} {414002} (\bibinfo {year} {2017})}\BibitemShut
  {NoStop}%
\bibitem [{\citenamefont {Bastianello}\ \emph {et~al.}(2018)\citenamefont
  {Bastianello}, \citenamefont {Piroli},\ and\ \citenamefont
  {Calabrese}}]{bastianello-18}%
  \BibitemOpen
  \bibfield  {author} {\bibinfo {author} {\bibfnamefont {A.}~\bibnamefont
  {Bastianello}}, \bibinfo {author} {\bibfnamefont {L.}~\bibnamefont {Piroli}},
  \ and\ \bibinfo {author} {\bibfnamefont {P.}~\bibnamefont {Calabrese}},\
  }\href {https://doi.org/10.1103/PhysRevLett.120.190601} {\bibfield  {journal}
  {\bibinfo  {journal} {Phys. Rev. Lett.}\ }\textbf {\bibinfo {volume} {120}},\
  \bibinfo {pages} {190601} (\bibinfo {year} {2018})}\BibitemShut {NoStop}%
\bibitem [{\citenamefont {Bastianello}\ and\ \citenamefont
  {Piroli}(2018)}]{bastianello-18-2}%
  \BibitemOpen
  \bibfield  {author} {\bibinfo {author} {\bibfnamefont {A.}~\bibnamefont
  {Bastianello}}\ and\ \bibinfo {author} {\bibfnamefont {L.}~\bibnamefont
  {Piroli}},\ }\href {https://doi.org/10.1088/1742-5468/aaeb48} {\bibfield
  {journal} {\bibinfo  {journal} {J. Stat. Mech.}\ ,\ \bibinfo {pages}
  {113104}} (\bibinfo {year} {2018})}\BibitemShut {NoStop}%
\bibitem [{\citenamefont {Arzamasovs}\ and\ \citenamefont
  {Gangardt}(2019)}]{arzamasovs-19}%
  \BibitemOpen
  \bibfield  {author} {\bibinfo {author} {\bibfnamefont {M.}~\bibnamefont
  {Arzamasovs}}\ and\ \bibinfo {author} {\bibfnamefont {D.}~\bibnamefont
  {Gangardt}},\ }\href {https://doi.org/10.1103/PhysRevLett.122.120401}
  {\bibfield  {journal} {\bibinfo  {journal} {Phys. Rev. Lett.}\ }\textbf
  {\bibinfo {volume} {122}},\ \bibinfo {pages} {120401} (\bibinfo {year}
  {2019})}\BibitemShut {NoStop}%
\bibitem [{\citenamefont {Najafi}\ and\ \citenamefont
  {Rajabpour}(2020)}]{najafi-20}%
  \BibitemOpen
  \bibfield  {author} {\bibinfo {author} {\bibfnamefont {M.~N.}\ \bibnamefont
  {Najafi}}\ and\ \bibinfo {author} {\bibfnamefont {M.~A.}\ \bibnamefont
  {Rajabpour}},\ }\href {https://doi.org/10.1103/PhysRevB.101.165415}
  {\bibfield  {journal} {\bibinfo  {journal} {Phys. Rev. B}\ }\textbf {\bibinfo
  {volume} {101}},\ \bibinfo {pages} {165415} (\bibinfo {year}
  {2020})}\BibitemShut {NoStop}%
\bibitem [{\citenamefont {Calabrese}\ \emph {et~al.}(2020)\citenamefont
  {Calabrese}, \citenamefont {Collura}, \citenamefont {Giulio},\ and\
  \citenamefont {Murciano}}]{calabrese-20}%
  \BibitemOpen
  \bibfield  {author} {\bibinfo {author} {\bibfnamefont {P.}~\bibnamefont
  {Calabrese}}, \bibinfo {author} {\bibfnamefont {M.}~\bibnamefont {Collura}},
  \bibinfo {author} {\bibfnamefont {G.~D.}\ \bibnamefont {Giulio}}, \ and\
  \bibinfo {author} {\bibfnamefont {S.}~\bibnamefont {Murciano}},\ }\href
  {https://doi.org/10.1209/0295-5075/129/60007} {\bibfield  {journal} {\bibinfo
   {journal} {EPL}\ }\textbf {\bibinfo {volume} {129}},\ \bibinfo {pages}
  {60007} (\bibinfo {year} {2020})}\BibitemShut {NoStop}%
\bibitem [{\citenamefont {Ares}\ \emph {et~al.}(2021)\citenamefont {Ares},
  \citenamefont {Rajabpour},\ and\ \citenamefont {Viti}}]{ares-21}%
  \BibitemOpen
  \bibfield  {author} {\bibinfo {author} {\bibfnamefont {F.}~\bibnamefont
  {Ares}}, \bibinfo {author} {\bibfnamefont {M.}~\bibnamefont {Rajabpour}}, \
  and\ \bibinfo {author} {\bibfnamefont {J.}~\bibnamefont {Viti}},\ }\href
  {https://doi.org/10.1103/PhysRevE.103.042107} {\bibfield  {journal} {\bibinfo
   {journal} {Phys. Rev. E}\ }\textbf {\bibinfo {volume} {103}},\ \bibinfo
  {pages} {042107} (\bibinfo {year} {2021})}\BibitemShut {NoStop}%
\bibitem [{\citenamefont {Cai}\ and\ \citenamefont {Cheng}()}]{cai-24}%
  \BibitemOpen
  \bibfield  {author} {\bibinfo {author} {\bibfnamefont {K.-L.}\ \bibnamefont
  {Cai}}\ and\ \bibinfo {author} {\bibfnamefont {M.}~\bibnamefont {Cheng}},\
  }\href {https://doi.org/10.48550/arXiv.2401.09548} {\bibinfo  {journal}
  {arXiv:2401.09548}\ }\BibitemShut {NoStop}%
\bibitem [{\citenamefont {Eisler}\ and\ \citenamefont
  {R\'acz}(2013)}]{eisler-13-2}%
  \BibitemOpen
\bibfield  {journal} {  }\bibfield  {author} {\bibinfo {author} {\bibfnamefont
  {V.}~\bibnamefont {Eisler}}\ and\ \bibinfo {author} {\bibfnamefont
  {Z.}~\bibnamefont {R\'acz}},\ }\href
  {https://doi.org/10.1103/PhysRevLett.110.060602} {\bibfield  {journal}
  {\bibinfo  {journal} {Phys. Rev. Lett.}\ }\textbf {\bibinfo {volume} {110}},\
  \bibinfo {pages} {060602} (\bibinfo {year} {2013})}\BibitemShut {NoStop}%
\bibitem [{\citenamefont {Lovas}\ \emph {et~al.}(2017)\citenamefont {Lovas},
  \citenamefont {D\'ora}, \citenamefont {Demler},\ and\ \citenamefont
  {Zar\'and}}]{lovas-17}%
  \BibitemOpen
  \bibfield  {author} {\bibinfo {author} {\bibfnamefont {I.}~\bibnamefont
  {Lovas}}, \bibinfo {author} {\bibfnamefont {B.}~\bibnamefont {D\'ora}},
  \bibinfo {author} {\bibfnamefont {E.}~\bibnamefont {Demler}}, \ and\ \bibinfo
  {author} {\bibfnamefont {G.}~\bibnamefont {Zar\'and}},\ }\href
  {https://doi.org/10.1103/PhysRevA.95.053621} {\bibfield  {journal} {\bibinfo
  {journal} {Phys. Rev. A}\ }\textbf {\bibinfo {volume} {95}},\ \bibinfo
  {pages} {053621} (\bibinfo {year} {2017})}\BibitemShut {NoStop}%
\bibitem [{\citenamefont {Groha}\ \emph {et~al.}(2018)\citenamefont {Groha},
  \citenamefont {Essler},\ and\ \citenamefont {Calabrese}}]{groha-18}%
  \BibitemOpen
  \bibfield  {author} {\bibinfo {author} {\bibfnamefont {S.}~\bibnamefont
  {Groha}}, \bibinfo {author} {\bibfnamefont {F.~H.~L.}\ \bibnamefont
  {Essler}}, \ and\ \bibinfo {author} {\bibfnamefont {P.}~\bibnamefont
  {Calabrese}},\ }\href {https://doi.org/10.21468/SciPostPhys.4.6.043}
  {\bibfield  {journal} {\bibinfo  {journal} {Sci. Post Phys.}\ }\textbf
  {\bibinfo {volume} {4}},\ \bibinfo {pages} {043} (\bibinfo {year}
  {2018})}\BibitemShut {NoStop}%
\bibitem [{\citenamefont {Collura}(2019)}]{collura-19}%
  \BibitemOpen
  \bibfield  {author} {\bibinfo {author} {\bibfnamefont {M.}~\bibnamefont
  {Collura}},\ }\href {https://doi.org/10.21468/SciPostPhys.7.6.072} {\bibfield
   {journal} {\bibinfo  {journal} {Sci. Post Phys.}\ }\textbf {\bibinfo
  {volume} {7}},\ \bibinfo {pages} {072} (\bibinfo {year} {2019})}\BibitemShut
  {NoStop}%
\bibitem [{\citenamefont {Gamayun}\ \emph {et~al.}(2020)\citenamefont
  {Gamayun}, \citenamefont {Lychkovskiy},\ and\ \citenamefont
  {Caux}}]{gamayun-19}%
  \BibitemOpen
  \bibfield  {author} {\bibinfo {author} {\bibfnamefont {O.}~\bibnamefont
  {Gamayun}}, \bibinfo {author} {\bibfnamefont {O.}~\bibnamefont
  {Lychkovskiy}}, \ and\ \bibinfo {author} {\bibfnamefont {J.-S.}\ \bibnamefont
  {Caux}},\ }\href {https://doi.org/10.21468/SciPostPhys.8.3.036} {\bibfield
  {journal} {\bibinfo  {journal} {Sci. Post Phys.}\ }\textbf {\bibinfo {volume}
  {8}},\ \bibinfo {pages} {036} (\bibinfo {year} {2020})}\BibitemShut {NoStop}%
\bibitem [{\citenamefont {Valencia-Tortora}\ \emph {et~al.}(2020)\citenamefont
  {Valencia-Tortora}, \citenamefont {Calabrese},\ and\ \citenamefont
  {Collura}}]{valencia-tortora-20}%
  \BibitemOpen
  \bibfield  {author} {\bibinfo {author} {\bibfnamefont {R.~J.}\ \bibnamefont
  {Valencia-Tortora}}, \bibinfo {author} {\bibfnamefont {P.}~\bibnamefont
  {Calabrese}}, \ and\ \bibinfo {author} {\bibfnamefont {M.}~\bibnamefont
  {Collura}},\ }\href
  {https://iopscience.iop.org/article/10.1209/0295-5075/132/50001/meta}
  {\bibfield  {journal} {\bibinfo  {journal} {EPL}\ }\textbf {\bibinfo {volume}
  {132}},\ \bibinfo {pages} {50001} (\bibinfo {year} {2020})}\BibitemShut
  {NoStop}%
\bibitem [{\citenamefont {Bertini}\ \emph {et~al.}(2023)\citenamefont
  {Bertini}, \citenamefont {Calabrese}, \citenamefont {Collura}, \citenamefont
  {Klobas},\ and\ \citenamefont {Rylands}}]{bertini-23}%
  \BibitemOpen
  \bibfield  {author} {\bibinfo {author} {\bibfnamefont {B.}~\bibnamefont
  {Bertini}}, \bibinfo {author} {\bibfnamefont {P.}~\bibnamefont {Calabrese}},
  \bibinfo {author} {\bibfnamefont {M.}~\bibnamefont {Collura}}, \bibinfo
  {author} {\bibfnamefont {K.}~\bibnamefont {Klobas}}, \ and\ \bibinfo {author}
  {\bibfnamefont {C.}~\bibnamefont {Rylands}},\ }\href
  {https://doi.org/10.1103/PhysRevLett.131.140401} {\bibfield  {journal}
  {\bibinfo  {journal} {Phys. Rev. Lett.}\ }\textbf {\bibinfo {volume} {131}},\
  \bibinfo {pages} {140401} (\bibinfo {year} {2023})}\BibitemShut {NoStop}%
\bibitem [{\citenamefont {Senese}\ \emph {et~al.}(2024)\citenamefont {Senese},
  \citenamefont {Robertson},\ and\ \citenamefont {Essler}}]{senese-23}%
  \BibitemOpen
  \bibfield  {author} {\bibinfo {author} {\bibfnamefont {R.}~\bibnamefont
  {Senese}}, \bibinfo {author} {\bibfnamefont {J.~H.}\ \bibnamefont
  {Robertson}}, \ and\ \bibinfo {author} {\bibfnamefont {F.~H.~L.}\
  \bibnamefont {Essler}},\ }\href {\doibase 10.21468/SciPostPhys.17.5.139}
  {\bibfield  {journal} {\bibinfo  {journal} {SciPost Phys.}\ }\textbf
  {\bibinfo {volume} {17}},\ \bibinfo {pages} {139} (\bibinfo {year}
  {2024})}\BibitemShut {NoStop}%
\bibitem [{\citenamefont {E.~Tirrito}\ and\ \citenamefont
  {Collura}(2023)}]{tirrito-23}%
  \BibitemOpen
  \bibfield  {author} {\bibinfo {author} {\bibfnamefont {R.~F.}\ \bibnamefont
  {E.~Tirrito}, \bibfnamefont {A.~Santini}}\ and\ \bibinfo {author}
  {\bibfnamefont {M.}~\bibnamefont {Collura}},\ }\href
  {https://www.scipost.org/10.21468/SciPostPhys.15.3.096} {\bibfield  {journal}
  {\bibinfo  {journal} {SciPost Phys.}\ }\textbf {\bibinfo {volume} {15}},\
  \bibinfo {pages} {096} (\bibinfo {year} {2023})}\BibitemShut {NoStop}%
\bibitem [{\citenamefont {Bertini}\ \emph {et~al.}(2024)\citenamefont
  {Bertini}, \citenamefont {Klobas}, \citenamefont {Collura}, \citenamefont
  {Calabrese},\ and\ \citenamefont {Rylands}}]{bertini-24}%
  \BibitemOpen
  \bibfield  {author} {\bibinfo {author} {\bibfnamefont {B.}~\bibnamefont
  {Bertini}}, \bibinfo {author} {\bibfnamefont {K.}~\bibnamefont {Klobas}},
  \bibinfo {author} {\bibfnamefont {M.}~\bibnamefont {Collura}}, \bibinfo
  {author} {\bibfnamefont {P.}~\bibnamefont {Calabrese}}, \ and\ \bibinfo
  {author} {\bibfnamefont {C.}~\bibnamefont {Rylands}},\ }\href
  {https://doi.org/10.1103/PhysRevB.109.184312} {\bibfield  {journal} {\bibinfo
   {journal} {Phys. Rev. B}\ }\textbf {\bibinfo {volume} {109}},\ \bibinfo
  {pages} {184312} (\bibinfo {year} {2024})}\BibitemShut {NoStop}%
\bibitem [{\citenamefont {Horvath}\ and\ \citenamefont
  {Rylands}(2024)}]{horvath-24}%
  \BibitemOpen
  \bibfield  {author} {\bibinfo {author} {\bibfnamefont {D.~X.}\ \bibnamefont
  {Horvath}}\ and\ \bibinfo {author} {\bibfnamefont {C.}~\bibnamefont
  {Rylands}},\ }\href {https://doi.org/10.1103/PhysRevA.109.043302} {\bibfield
  {journal} {\bibinfo  {journal} {Phys. Rev. A}\ }\textbf {\bibinfo {volume}
  {109}},\ \bibinfo {pages} {043302} (\bibinfo {year} {2024})}\BibitemShut
  {NoStop}%
\bibitem [{\citenamefont {Horvath}\ \emph {et~al.}()\citenamefont {Horvath},
  \citenamefont {Doyon},\ and\ \citenamefont {Ruggiero}}]{horvath-24-2}%
  \BibitemOpen
  \bibfield  {author} {\bibinfo {author} {\bibfnamefont {D.~X.}\ \bibnamefont
  {Horvath}}, \bibinfo {author} {\bibfnamefont {B.}~\bibnamefont {Doyon}}, \
  and\ \bibinfo {author} {\bibfnamefont {P.}~\bibnamefont {Ruggiero}},\ }\href
  {https://doi.org/10.48550/arXiv.2411.14406} {\bibinfo  {journal}
  {arXiv:2411.14406}\ }\BibitemShut {NoStop}%
\bibitem [{\citenamefont {Ranabhat}\ and\ \citenamefont
  {Collura}(2024)}]{ranabhat-24}%
  \BibitemOpen
\bibfield  {journal} {  }\bibfield  {author} {\bibinfo {author} {\bibfnamefont
  {N.}~\bibnamefont {Ranabhat}}\ and\ \bibinfo {author} {\bibfnamefont
  {M.}~\bibnamefont {Collura}},\ }\href
  {https://scipost.org/SciPostPhysCore.7.2.017} {\bibfield  {journal} {\bibinfo
   {journal} {SciPost Phys. Core}\ }\textbf {\bibinfo {volume} {7}},\ \bibinfo
  {pages} {017} (\bibinfo {year} {2024})}\BibitemShut {NoStop}%
\bibitem [{\citenamefont {Klich}\ and\ \citenamefont
  {Levitov}(2009{\natexlab{a}})}]{klich-09}%
  \BibitemOpen
  \bibfield  {author} {\bibinfo {author} {\bibfnamefont {I.}~\bibnamefont
  {Klich}}\ and\ \bibinfo {author} {\bibfnamefont {L.}~\bibnamefont
  {Levitov}},\ }\href {https://doi.org/10.1103/PhysRevLett.102.100502}
  {\bibfield  {journal} {\bibinfo  {journal} {Phys. Rev. Lett.}\ }\textbf
  {\bibinfo {volume} {102}},\ \bibinfo {pages} {100502} (\bibinfo {year}
  {2009}{\natexlab{a}})}\BibitemShut {NoStop}%
\bibitem [{\citenamefont {Klich}\ and\ \citenamefont
  {Levitov}(2009{\natexlab{b}})}]{klich-09-2}%
  \BibitemOpen
  \bibfield  {author} {\bibinfo {author} {\bibfnamefont {I.}~\bibnamefont
  {Klich}}\ and\ \bibinfo {author} {\bibfnamefont {L.}~\bibnamefont
  {Levitov}},\ }\href {https://doi.org/10.1063/1.3149497} {\bibfield  {journal}
  {\bibinfo  {journal} {Adv. Theor. Phys.}\ }\textbf {\bibinfo {volume}
  {1134}},\ \bibinfo {pages} {36} (\bibinfo {year}
  {2009}{\natexlab{b}})}\BibitemShut {NoStop}%
\bibitem [{\citenamefont {Song}\ \emph {et~al.}(2011)\citenamefont {Song},
  \citenamefont {Flindt}, \citenamefont {Rachel}, \citenamefont {Klich},\ and\
  \citenamefont {Hur}}]{song-11}%
  \BibitemOpen
  \bibfield  {author} {\bibinfo {author} {\bibfnamefont {H.~F.}\ \bibnamefont
  {Song}}, \bibinfo {author} {\bibfnamefont {C.}~\bibnamefont {Flindt}},
  \bibinfo {author} {\bibfnamefont {S.}~\bibnamefont {Rachel}}, \bibinfo
  {author} {\bibfnamefont {I.}~\bibnamefont {Klich}}, \ and\ \bibinfo {author}
  {\bibfnamefont {K.~L.}\ \bibnamefont {Hur}},\ }\href
  {http://dx.doi.org/10.1103/PhysRevB.83.161408} {\bibfield  {journal}
  {\bibinfo  {journal} {Phys. Rev. B}\ }\textbf {\bibinfo {volume} {83}},\
  \bibinfo {pages} {161408(R)} (\bibinfo {year} {2011})}\BibitemShut {NoStop}%
\bibitem [{\citenamefont {Song}\ \emph {et~al.}(2012)\citenamefont {Song},
  \citenamefont {Rachel}, \citenamefont {Flindt}, \citenamefont {Klich},
  \citenamefont {Laflorencie},\ and\ \citenamefont {Hur}}]{song-12}%
  \BibitemOpen
  \bibfield  {author} {\bibinfo {author} {\bibfnamefont {H.~F.}\ \bibnamefont
  {Song}}, \bibinfo {author} {\bibfnamefont {S.}~\bibnamefont {Rachel}},
  \bibinfo {author} {\bibfnamefont {C.}~\bibnamefont {Flindt}}, \bibinfo
  {author} {\bibfnamefont {I.}~\bibnamefont {Klich}}, \bibinfo {author}
  {\bibfnamefont {N.}~\bibnamefont {Laflorencie}}, \ and\ \bibinfo {author}
  {\bibfnamefont {K.~L.}\ \bibnamefont {Hur}},\ }\href
  {http://dx.doi.org/10.1103/PhysRevB.85.035409} {\bibfield  {journal}
  {\bibinfo  {journal} {Phys. Rev. B}\ }\textbf {\bibinfo {volume} {85}},\
  \bibinfo {pages} {035409} (\bibinfo {year} {2012})}\BibitemShut {NoStop}%
\bibitem [{\citenamefont {Calabrese}\ \emph {et~al.}(2012)\citenamefont
  {Calabrese}, \citenamefont {Mintchev},\ and\ \citenamefont
  {Vicari}}]{calabrese-12}%
  \BibitemOpen
  \bibfield  {author} {\bibinfo {author} {\bibfnamefont {P.}~\bibnamefont
  {Calabrese}}, \bibinfo {author} {\bibfnamefont {M.}~\bibnamefont {Mintchev}},
  \ and\ \bibinfo {author} {\bibfnamefont {E.}~\bibnamefont {Vicari}},\ }\href
  {http://dx.doi.org/10.1209/0295-5075/98/20003} {\bibfield  {journal}
  {\bibinfo  {journal} {EPL}\ }\textbf {\bibinfo {volume} {98}},\ \bibinfo
  {pages} {20003} (\bibinfo {year} {2012})}\BibitemShut {NoStop}%
\bibitem [{\citenamefont {Levine}\ \emph {et~al.}(2012)\citenamefont {Levine},
  \citenamefont {Bantegui},\ and\ \citenamefont {Burg}}]{levine-12}%
  \BibitemOpen
  \bibfield  {author} {\bibinfo {author} {\bibfnamefont {G.~C.}\ \bibnamefont
  {Levine}}, \bibinfo {author} {\bibfnamefont {M.~J.}\ \bibnamefont
  {Bantegui}}, \ and\ \bibinfo {author} {\bibfnamefont {J.~A.}\ \bibnamefont
  {Burg}},\ }\href {https://doi.org/10.1103/PhysRevB.86.174202} {\bibfield
  {journal} {\bibinfo  {journal} {Phys. Rev. B}\ }\textbf {\bibinfo {volume}
  {86}},\ \bibinfo {pages} {174202} (\bibinfo {year} {2012})}\BibitemShut
  {NoStop}%
\bibitem [{\citenamefont {Susstrunk}\ and\ \citenamefont
  {Ivanov}(2012)}]{susstrunk-12}%
  \BibitemOpen
  \bibfield  {author} {\bibinfo {author} {\bibfnamefont {R.}~\bibnamefont
  {Susstrunk}}\ and\ \bibinfo {author} {\bibfnamefont {D.~A.}\ \bibnamefont
  {Ivanov}},\ }\href {https://doi.org/10.1209/0295-5075/100/60009} {\bibfield
  {journal} {\bibinfo  {journal} {EPL}\ }\textbf {\bibinfo {volume} {100}},\
  \bibinfo {pages} {60009} (\bibinfo {year} {2012})}\BibitemShut {NoStop}%
\bibitem [{\citenamefont {Calabrese}\ \emph {et~al.}(2015)\citenamefont
  {Calabrese}, \citenamefont {Doussal},\ and\ \citenamefont
  {Majumdar}}]{calabrese-15}%
  \BibitemOpen
  \bibfield  {author} {\bibinfo {author} {\bibfnamefont {P.}~\bibnamefont
  {Calabrese}}, \bibinfo {author} {\bibfnamefont {P.~L.}\ \bibnamefont
  {Doussal}}, \ and\ \bibinfo {author} {\bibfnamefont {S.~N.}\ \bibnamefont
  {Majumdar}},\ }\href {https://doi.org/10.1103/PhysRevA.91.012303} {\bibfield
  {journal} {\bibinfo  {journal} {Phys. Rev. A}\ }\textbf {\bibinfo {volume}
  {91}},\ \bibinfo {pages} {012303} (\bibinfo {year} {2015})}\BibitemShut
  {NoStop}%
\bibitem [{\citenamefont {Utsumi}(2019)}]{utsumi-19}%
  \BibitemOpen
  \bibfield  {author} {\bibinfo {author} {\bibfnamefont {Y.}~\bibnamefont
  {Utsumi}},\ }\href {https://doi.org/10.1140/epjst/e2018-800043-4} {\bibfield
  {journal} {\bibinfo  {journal} {Eur. Phys. J. Spec. Top.}\ }\textbf {\bibinfo
  {volume} {227}},\ \bibinfo {pages} {1911} (\bibinfo {year}
  {2019})}\BibitemShut {NoStop}%
\bibitem [{\citenamefont {Goldstein}\ and\ \citenamefont
  {Sela}(2018)}]{goldstein-18}%
  \BibitemOpen
  \bibfield  {author} {\bibinfo {author} {\bibfnamefont {M.}~\bibnamefont
  {Goldstein}}\ and\ \bibinfo {author} {\bibfnamefont {E.}~\bibnamefont
  {Sela}},\ }\href {https://doi.org/10.1103/PhysRevLett.120.200602} {\bibfield
  {journal} {\bibinfo  {journal} {Phys. Rev. Lett.}\ }\textbf {\bibinfo
  {volume} {120}},\ \bibinfo {pages} {200602} (\bibinfo {year}
  {2018})}\BibitemShut {NoStop}%
\bibitem [{\citenamefont {Xavier}\ \emph {et~al.}(2018)\citenamefont {Xavier},
  \citenamefont {Alcaraz},\ and\ \citenamefont {Sierra}}]{xavier-18}%
  \BibitemOpen
  \bibfield  {author} {\bibinfo {author} {\bibfnamefont {J.~C.}\ \bibnamefont
  {Xavier}}, \bibinfo {author} {\bibfnamefont {F.~C.}\ \bibnamefont {Alcaraz}},
  \ and\ \bibinfo {author} {\bibfnamefont {G.}~\bibnamefont {Sierra}},\ }\href
  {https://doi.org/10.1103/PhysRevB.98.041106} {\bibfield  {journal} {\bibinfo
  {journal} {Phys. Rev. B}\ }\textbf {\bibinfo {volume} {98}},\ \bibinfo
  {pages} {041106} (\bibinfo {year} {2018})}\BibitemShut {NoStop}%
\bibitem [{\citenamefont {Lukin}\ \emph {et~al.}(2019)\citenamefont {Lukin},
  \citenamefont {Rispoli}, \citenamefont {Schittko}, \citenamefont {Tai},
  \citenamefont {Kaufman}, \citenamefont {Choi}, \citenamefont {Khemani},
  \citenamefont {Leonard},\ and\ \citenamefont {Greiner}}]{lukin-19}%
  \BibitemOpen
  \bibfield  {author} {\bibinfo {author} {\bibfnamefont {A.}~\bibnamefont
  {Lukin}}, \bibinfo {author} {\bibfnamefont {M.}~\bibnamefont {Rispoli}},
  \bibinfo {author} {\bibfnamefont {R.}~\bibnamefont {Schittko}}, \bibinfo
  {author} {\bibfnamefont {M.~E.}\ \bibnamefont {Tai}}, \bibinfo {author}
  {\bibfnamefont {A.~M.}\ \bibnamefont {Kaufman}}, \bibinfo {author}
  {\bibfnamefont {S.}~\bibnamefont {Choi}}, \bibinfo {author} {\bibfnamefont
  {V.}~\bibnamefont {Khemani}}, \bibinfo {author} {\bibfnamefont
  {J.}~\bibnamefont {Leonard}}, \ and\ \bibinfo {author} {\bibfnamefont
  {M.}~\bibnamefont {Greiner}},\ }\href
  {https://dx.doi.org/10.1126/science.aau0818} {\bibfield  {journal} {\bibinfo
  {journal} {Science}\ }\textbf {\bibinfo {volume} {364}},\ \bibinfo {pages}
  {6437} (\bibinfo {year} {2019})}\BibitemShut {NoStop}%
\bibitem [{\citenamefont {Ares}\ \emph {et~al.}(2023)\citenamefont {Ares},
  \citenamefont {Murciano},\ and\ \citenamefont {Calabrese}}]{ares-23}%
  \BibitemOpen
  \bibfield  {author} {\bibinfo {author} {\bibfnamefont {F.}~\bibnamefont
  {Ares}}, \bibinfo {author} {\bibfnamefont {S.}~\bibnamefont {Murciano}}, \
  and\ \bibinfo {author} {\bibfnamefont {P.}~\bibnamefont {Calabrese}},\ }\href
  {https://doi.org/10.1038/s41467-023-37747-8} {\bibfield  {journal} {\bibinfo
  {journal} {Nature Comms.}\ }\textbf {\bibinfo {volume} {14}},\ \bibinfo
  {pages} {2036} (\bibinfo {year} {2023})}\BibitemShut {NoStop}%
\bibitem [{\citenamefont {Rylands}\ \emph {et~al.}(2024)\citenamefont
  {Rylands}, \citenamefont {Klobas}, \citenamefont {Ares}, \citenamefont
  {Calabrese}, \citenamefont {Murciano},\ and\ \citenamefont
  {Bertini}}]{rylands-24}%
  \BibitemOpen
  \bibfield  {author} {\bibinfo {author} {\bibfnamefont {C.}~\bibnamefont
  {Rylands}}, \bibinfo {author} {\bibfnamefont {K.}~\bibnamefont {Klobas}},
  \bibinfo {author} {\bibfnamefont {F.}~\bibnamefont {Ares}}, \bibinfo {author}
  {\bibfnamefont {P.}~\bibnamefont {Calabrese}}, \bibinfo {author}
  {\bibfnamefont {S.}~\bibnamefont {Murciano}}, \ and\ \bibinfo {author}
  {\bibfnamefont {B.}~\bibnamefont {Bertini}},\ }\href
  {https://doi.org/10.1103/PhysRevLett.133.010401} {\bibfield  {journal}
  {\bibinfo  {journal} {Phys. Rev. Lett.}\ }\textbf {\bibinfo {volume} {133}},\
  \bibinfo {pages} {010401} (\bibinfo {year} {2024})}\BibitemShut {NoStop}%
\bibitem [{\citenamefont {Wienand}\ \emph {et~al.}(2024)\citenamefont
  {Wienand}, \citenamefont {Karch}, \citenamefont {Impertro}, \citenamefont
  {Schweizer}, \citenamefont {McCulloch}, \citenamefont {Vasseur},
  \citenamefont {Gopalakrishnan}, \citenamefont {Aidelsburger},\ and\
  \citenamefont {Bloch}}]{Wienand2024}%
  \BibitemOpen
  \bibfield  {author} {\bibinfo {author} {\bibfnamefont {J.~F.}\ \bibnamefont
  {Wienand}}, \bibinfo {author} {\bibfnamefont {S.}~\bibnamefont {Karch}},
  \bibinfo {author} {\bibfnamefont {A.}~\bibnamefont {Impertro}}, \bibinfo
  {author} {\bibfnamefont {C.}~\bibnamefont {Schweizer}}, \bibinfo {author}
  {\bibfnamefont {E.}~\bibnamefont {McCulloch}}, \bibinfo {author}
  {\bibfnamefont {R.}~\bibnamefont {Vasseur}}, \bibinfo {author} {\bibfnamefont
  {S.}~\bibnamefont {Gopalakrishnan}}, \bibinfo {author} {\bibfnamefont
  {M.}~\bibnamefont {Aidelsburger}}, \ and\ \bibinfo {author} {\bibfnamefont
  {I.}~\bibnamefont {Bloch}},\ }\href
  {https://doi.org/10.1038/s41567-024-02611-z} {\bibfield  {journal} {\bibinfo
  {journal} {Nature Physics}\ } (\bibinfo {year} {2024})}\BibitemShut {NoStop}%
\bibitem [{\citenamefont {{Google AI and collaborators}}(2024)}]{google-24}%
  \BibitemOpen
  \bibfield  {author} {\bibinfo {author} {\bibnamefont {{Google AI and
  collaborators}}},\ }\href {https://doi.org/10.1126/science.adi7877}
  {\bibfield  {journal} {\bibinfo  {journal} {Science}\ }\textbf {\bibinfo
  {volume} {384}},\ \bibinfo {pages} {48} (\bibinfo {year} {2024})}\BibitemShut
  {NoStop}%
\bibitem [{\citenamefont {Wei}\ \emph {et~al.}(2022)\citenamefont {Wei},
  \citenamefont {Rubio-Abadal}, \citenamefont {Ye}, \citenamefont {Machado},
  \citenamefont {Kemp}, \citenamefont {Srakaew}, \citenamefont {Hollerith},
  \citenamefont {Rui}, \citenamefont {Gopalakrishnan}, \citenamefont {Yao},
  \citenamefont {Bloch},\ and\ \citenamefont {Zeihe}}]{wei-22}%
  \BibitemOpen
  \bibfield  {author} {\bibinfo {author} {\bibfnamefont {D.}~\bibnamefont
  {Wei}}, \bibinfo {author} {\bibfnamefont {A.}~\bibnamefont {Rubio-Abadal}},
  \bibinfo {author} {\bibfnamefont {B.}~\bibnamefont {Ye}}, \bibinfo {author}
  {\bibfnamefont {F.}~\bibnamefont {Machado}}, \bibinfo {author} {\bibfnamefont
  {J.}~\bibnamefont {Kemp}}, \bibinfo {author} {\bibfnamefont {K.}~\bibnamefont
  {Srakaew}}, \bibinfo {author} {\bibfnamefont {S.}~\bibnamefont {Hollerith}},
  \bibinfo {author} {\bibfnamefont {J.}~\bibnamefont {Rui}}, \bibinfo {author}
  {\bibfnamefont {S.}~\bibnamefont {Gopalakrishnan}}, \bibinfo {author}
  {\bibfnamefont {N.~Y.}\ \bibnamefont {Yao}}, \bibinfo {author} {\bibfnamefont
  {I.}~\bibnamefont {Bloch}}, \ and\ \bibinfo {author} {\bibfnamefont
  {J.}~\bibnamefont {Zeihe}},\ }\href {https://doi.org/10.1126/science.abk2397}
  {\bibfield  {journal} {\bibinfo  {journal} {Science}\ }\textbf {\bibinfo
  {volume} {376}},\ \bibinfo {pages} {716} (\bibinfo {year}
  {2022})}\BibitemShut {NoStop}%
\bibitem [{\citenamefont {Valli}\ \emph {et~al.}()\citenamefont {Valli},
  \citenamefont {Moca}, \citenamefont {Werner}, \citenamefont {Kormos},
  \citenamefont {Krajnik}, \citenamefont {Prosen},\ and\ \citenamefont
  {Zar\'and}}]{valli-24}%
  \BibitemOpen
  \bibfield  {author} {\bibinfo {author} {\bibfnamefont {A.}~\bibnamefont
  {Valli}}, \bibinfo {author} {\bibfnamefont {C.~P.}\ \bibnamefont {Moca}},
  \bibinfo {author} {\bibfnamefont {M.~A.}\ \bibnamefont {Werner}}, \bibinfo
  {author} {\bibfnamefont {M.}~\bibnamefont {Kormos}}, \bibinfo {author}
  {\bibfnamefont {Z.}~\bibnamefont {Krajnik}}, \bibinfo {author} {\bibfnamefont
  {T.}~\bibnamefont {Prosen}}, \ and\ \bibinfo {author} {\bibfnamefont
  {G.}~\bibnamefont {Zar\'and}},\ }\href
  {https://doi.org/10.48550/arXiv.2409.14442} {\bibinfo  {journal}
  {arXiv:2409.14442}\ }\BibitemShut {NoStop}%
\bibitem [{\citenamefont {Brydges}\ \emph {et~al.}(2019)\citenamefont
  {Brydges}, \citenamefont {Elben}, \citenamefont {Jurcevic}, \citenamefont
  {Vermersch}, \citenamefont {Maier}, \citenamefont {Lanyon}, \citenamefont
  {Zoller}, \citenamefont {Blatt},\ and\ \citenamefont {Roos}}]{Brydges2019}%
  \BibitemOpen
\bibfield  {journal} {  }\bibfield  {author} {\bibinfo {author} {\bibfnamefont
  {T.}~\bibnamefont {Brydges}}, \bibinfo {author} {\bibfnamefont
  {A.}~\bibnamefont {Elben}}, \bibinfo {author} {\bibfnamefont
  {P.}~\bibnamefont {Jurcevic}}, \bibinfo {author} {\bibfnamefont
  {B.}~\bibnamefont {Vermersch}}, \bibinfo {author} {\bibfnamefont
  {C.}~\bibnamefont {Maier}}, \bibinfo {author} {\bibfnamefont {B.~P.}\
  \bibnamefont {Lanyon}}, \bibinfo {author} {\bibfnamefont {P.}~\bibnamefont
  {Zoller}}, \bibinfo {author} {\bibfnamefont {R.}~\bibnamefont {Blatt}}, \
  and\ \bibinfo {author} {\bibfnamefont {C.~F.}\ \bibnamefont {Roos}},\ }\href
  {\doibase 10.1126/science.aau4963} {\bibfield  {journal} {\bibinfo  {journal}
  {Science}\ }\textbf {\bibinfo {volume} {364}},\ \bibinfo {pages} {260}
  (\bibinfo {year} {2019})}\BibitemShut {NoStop}%
\bibitem [{\citenamefont {Joshi}\ \emph {et~al.}(2024)\citenamefont {Joshi},
  \citenamefont {Franke}, \citenamefont {Rath}, \citenamefont {Ares},
  \citenamefont {Murciano}, \citenamefont {Kranzl}, \citenamefont {Blatt},
  \citenamefont {Zoller}, \citenamefont {Vermersch}, \citenamefont {Calabrese},
  \citenamefont {Roos},\ and\ \citenamefont {Joshi}}]{joshi-24}%
  \BibitemOpen
  \bibfield  {author} {\bibinfo {author} {\bibfnamefont {L.~K.}\ \bibnamefont
  {Joshi}}, \bibinfo {author} {\bibfnamefont {J.}~\bibnamefont {Franke}},
  \bibinfo {author} {\bibfnamefont {A.}~\bibnamefont {Rath}}, \bibinfo {author}
  {\bibfnamefont {F.}~\bibnamefont {Ares}}, \bibinfo {author} {\bibfnamefont
  {S.}~\bibnamefont {Murciano}}, \bibinfo {author} {\bibfnamefont
  {F.}~\bibnamefont {Kranzl}}, \bibinfo {author} {\bibfnamefont
  {R.}~\bibnamefont {Blatt}}, \bibinfo {author} {\bibfnamefont
  {P.}~\bibnamefont {Zoller}}, \bibinfo {author} {\bibfnamefont
  {B.}~\bibnamefont {Vermersch}}, \bibinfo {author} {\bibfnamefont
  {P.}~\bibnamefont {Calabrese}}, \bibinfo {author} {\bibfnamefont {C.~F.}\
  \bibnamefont {Roos}}, \ and\ \bibinfo {author} {\bibfnamefont {M.~K.}\
  \bibnamefont {Joshi}},\ }\href {\doibase 10.1103/PhysRevLett.133.010402}
  {\bibfield  {journal} {\bibinfo  {journal} {Phys. Rev. Lett.}\ }\textbf
  {\bibinfo {volume} {133}},\ \bibinfo {pages} {010402} (\bibinfo {year}
  {2024})}\BibitemShut {NoStop}%
\bibitem [{LKJ()}]{LKJSM_FCS2025}%
  \BibitemOpen
  \href@noop {} {\bibinfo  {journal} {See Supplemental Material (SM) for
  further technical details. The SM includes the reference \cite{Mezzadri2006}
  detailing the generation of random unitaries}\ }\BibitemShut {NoStop}%
\bibitem [{\citenamefont {Huang}\ \emph {et~al.}(2020)\citenamefont {Huang},
  \citenamefont {Kueng},\ and\ \citenamefont {Preskill}}]{Huang2020}%
  \BibitemOpen
\bibfield  {journal} {  }\bibfield  {author} {\bibinfo {author} {\bibfnamefont
  {H.-Y.}\ \bibnamefont {Huang}}, \bibinfo {author} {\bibfnamefont
  {R.}~\bibnamefont {Kueng}}, \ and\ \bibinfo {author} {\bibfnamefont
  {J.}~\bibnamefont {Preskill}},\ }\href {\doibase 10.1038/s41567-020-0932-7}
  {\bibfield  {journal} {\bibinfo  {journal} {Nat. Phys.}\ }\textbf {\bibinfo
  {volume} {16}},\ \bibinfo {pages} {1050} (\bibinfo {year}
  {2020})}\BibitemShut {NoStop}%
\bibitem [{\citenamefont {Elben}\ \emph {et~al.}(2023)\citenamefont {Elben},
  \citenamefont {Flammia}, \citenamefont {Huang}, \citenamefont {Kueng},
  \citenamefont {Preskill}, \citenamefont {Vermersch},\ and\ \citenamefont
  {Zoller}}]{elben_review}%
  \BibitemOpen
  \bibfield  {author} {\bibinfo {author} {\bibfnamefont {A.}~\bibnamefont
  {Elben}}, \bibinfo {author} {\bibfnamefont {S.~T.}\ \bibnamefont {Flammia}},
  \bibinfo {author} {\bibfnamefont {H.-Y.}\ \bibnamefont {Huang}}, \bibinfo
  {author} {\bibfnamefont {R.}~\bibnamefont {Kueng}}, \bibinfo {author}
  {\bibfnamefont {J.}~\bibnamefont {Preskill}}, \bibinfo {author}
  {\bibfnamefont {B.}~\bibnamefont {Vermersch}}, \ and\ \bibinfo {author}
  {\bibfnamefont {P.}~\bibnamefont {Zoller}},\ }\href {\doibase
  10.1038/s42254-022-00535-2} {\bibfield  {journal} {\bibinfo  {journal} {Nat.
  Rev. Phys.}\ }\textbf {\bibinfo {volume} {5}},\ \bibinfo {pages} {9}
  (\bibinfo {year} {2023})}\BibitemShut {NoStop}%
\bibitem [{\citenamefont {Rath}\ \emph {et~al.}(2021)\citenamefont {Rath},
  \citenamefont {van Bijnen}, \citenamefont {Elben}, \citenamefont {Zoller},\
  and\ \citenamefont {Vermersch}}]{Rath2021}%
  \BibitemOpen
  \bibfield  {author} {\bibinfo {author} {\bibfnamefont {A.}~\bibnamefont
  {Rath}}, \bibinfo {author} {\bibfnamefont {R.}~\bibnamefont {van Bijnen}},
  \bibinfo {author} {\bibfnamefont {A.}~\bibnamefont {Elben}}, \bibinfo
  {author} {\bibfnamefont {P.}~\bibnamefont {Zoller}}, \ and\ \bibinfo {author}
  {\bibfnamefont {B.}~\bibnamefont {Vermersch}},\ }\href {\doibase
  10.1103/PhysRevLett.127.200503} {\bibfield  {journal} {\bibinfo  {journal}
  {Phys. Rev. Lett.}\ }\textbf {\bibinfo {volume} {127}},\ \bibinfo {pages}
  {200503} (\bibinfo {year} {2021})}\BibitemShut {NoStop}%
\bibitem [{\citenamefont {Satzinger}\ \emph {et~al.}(2021)\citenamefont
  {Satzinger}, \citenamefont {Liu}, \citenamefont {Smith}, \citenamefont
  {Knapp} \emph {et~al.}}]{Satzinger2021}%
  \BibitemOpen
  \bibfield  {author} {\bibinfo {author} {\bibfnamefont {K.~J.}\ \bibnamefont
  {Satzinger}}, \bibinfo {author} {\bibfnamefont {Y.-J.}\ \bibnamefont {Liu}},
  \bibinfo {author} {\bibfnamefont {A.}~\bibnamefont {Smith}}, \bibinfo
  {author} {\bibfnamefont {C.}~\bibnamefont {Knapp}},  \emph {et~al.},\ }\href
  {\doibase 10.1126/science.abi8378} {\bibfield  {journal} {\bibinfo  {journal}
  {Science}\ }\textbf {\bibinfo {volume} {374}},\ \bibinfo {pages} {1237}
  (\bibinfo {year} {2021})}\BibitemShut {NoStop}%
\bibitem [{\citenamefont {Hoke}\ \emph {et~al.}(2023)\citenamefont {Hoke},
  \citenamefont {Ippoliti}, \citenamefont {Rosenberg} \emph
  {et~al.}}]{Hoke2023}%
  \BibitemOpen
  \bibfield  {author} {\bibinfo {author} {\bibfnamefont {J.~C.}\ \bibnamefont
  {Hoke}}, \bibinfo {author} {\bibfnamefont {M.}~\bibnamefont {Ippoliti}},
  \bibinfo {author} {\bibfnamefont {E.}~\bibnamefont {Rosenberg}},  \emph
  {et~al.},\ }\href {\doibase 10.1038/s41586-023-06505-7} {\bibfield  {journal}
  {\bibinfo  {journal} {Nature}\ }\textbf {\bibinfo {volume} {622}},\ \bibinfo
  {pages} {481} (\bibinfo {year} {2023})}\BibitemShut {NoStop}%
\bibitem [{\citenamefont {Rath}\ \emph {et~al.}(2023)\citenamefont {Rath},
  \citenamefont {Vitale}, \citenamefont {Murciano} \emph {et~al.}}]{rath-23}%
  \BibitemOpen
  \bibfield  {author} {\bibinfo {author} {\bibfnamefont {A.}~\bibnamefont
  {Rath}}, \bibinfo {author} {\bibfnamefont {V.}~\bibnamefont {Vitale}},
  \bibinfo {author} {\bibfnamefont {S.}~\bibnamefont {Murciano}},  \emph
  {et~al.},\ }\href {https://doi.org/10.1103/PRXQuantum.4.010318} {\bibfield
  {journal} {\bibinfo  {journal} {PRX Quantum}\ }\textbf {\bibinfo {volume}
  {4}},\ \bibinfo {pages} {010318} (\bibinfo {year} {2023})}\BibitemShut
  {NoStop}%
\bibitem [{\citenamefont {Zhou}\ \emph {et~al.}(2020)\citenamefont {Zhou},
  \citenamefont {Zeng},\ and\ \citenamefont {Liu}}]{Zhou2020}%
  \BibitemOpen
  \bibfield  {author} {\bibinfo {author} {\bibfnamefont {Y.}~\bibnamefont
  {Zhou}}, \bibinfo {author} {\bibfnamefont {P.}~\bibnamefont {Zeng}}, \ and\
  \bibinfo {author} {\bibfnamefont {Z.}~\bibnamefont {Liu}},\ }\href
  {https://link.aps.org/doi/10.1103/PhysRevLett.125.200502} {\bibfield
  {journal} {\bibinfo  {journal} {Phys. Rev. Lett.}\ }\textbf {\bibinfo
  {volume} {125}},\ \bibinfo {pages} {200502} (\bibinfo {year}
  {2020})}\BibitemShut {NoStop}%
\bibitem [{\citenamefont {Elben}\ \emph {et~al.}(2020)\citenamefont {Elben},
  \citenamefont {Kueng}, \citenamefont {Huang}, \citenamefont {van Bijnen},
  \citenamefont {Kokail}, \citenamefont {Dalmonte}, \citenamefont {Calabrese},
  \citenamefont {Kraus}, \citenamefont {Preskill}, \citenamefont {Zoller},\
  and\ \citenamefont {Vermersch}}]{Elben2020b}%
  \BibitemOpen
  \bibfield  {author} {\bibinfo {author} {\bibfnamefont {A.}~\bibnamefont
  {Elben}}, \bibinfo {author} {\bibfnamefont {R.}~\bibnamefont {Kueng}},
  \bibinfo {author} {\bibfnamefont {H.-Y.~R.}\ \bibnamefont {Huang}}, \bibinfo
  {author} {\bibfnamefont {R.}~\bibnamefont {van Bijnen}}, \bibinfo {author}
  {\bibfnamefont {C.}~\bibnamefont {Kokail}}, \bibinfo {author} {\bibfnamefont
  {M.}~\bibnamefont {Dalmonte}}, \bibinfo {author} {\bibfnamefont
  {P.}~\bibnamefont {Calabrese}}, \bibinfo {author} {\bibfnamefont
  {B.}~\bibnamefont {Kraus}}, \bibinfo {author} {\bibfnamefont
  {J.}~\bibnamefont {Preskill}}, \bibinfo {author} {\bibfnamefont
  {P.}~\bibnamefont {Zoller}}, \ and\ \bibinfo {author} {\bibfnamefont
  {B.}~\bibnamefont {Vermersch}},\ }\href
  {https://link.aps.org/doi/10.1103/PhysRevLett.125.200501} {\bibfield
  {journal} {\bibinfo  {journal} {Phys. Rev. Lett.}\ }\textbf {\bibinfo
  {volume} {125}},\ \bibinfo {pages} {200501} (\bibinfo {year}
  {2020})}\BibitemShut {NoStop}%
\bibitem [{\citenamefont {Neven}\ \emph {et~al.}(2021)\citenamefont {Neven},
  \citenamefont {Carrasco}, \citenamefont {Vitale}, \citenamefont {Kokail},
  \citenamefont {Elben}, \citenamefont {Dalmonte}, \citenamefont {Calabrese},
  \citenamefont {Zoller}, \citenamefont {Vermersch}, \citenamefont {Kueng},\
  and\ \citenamefont {Kraus}}]{Neven2021}%
  \BibitemOpen
  \bibfield  {author} {\bibinfo {author} {\bibfnamefont {A.}~\bibnamefont
  {Neven}}, \bibinfo {author} {\bibfnamefont {J.}~\bibnamefont {Carrasco}},
  \bibinfo {author} {\bibfnamefont {V.}~\bibnamefont {Vitale}}, \bibinfo
  {author} {\bibfnamefont {C.}~\bibnamefont {Kokail}}, \bibinfo {author}
  {\bibfnamefont {A.}~\bibnamefont {Elben}}, \bibinfo {author} {\bibfnamefont
  {M.}~\bibnamefont {Dalmonte}}, \bibinfo {author} {\bibfnamefont
  {P.}~\bibnamefont {Calabrese}}, \bibinfo {author} {\bibfnamefont
  {P.}~\bibnamefont {Zoller}}, \bibinfo {author} {\bibfnamefont
  {B.}~\bibnamefont {Vermersch}}, \bibinfo {author} {\bibfnamefont
  {R.}~\bibnamefont {Kueng}}, \ and\ \bibinfo {author} {\bibfnamefont
  {B.}~\bibnamefont {Kraus}},\ }\href
  {https://www.nature.com/articles/s41534-021-00487-y} {\bibfield  {journal}
  {\bibinfo  {journal} {npj Quantum Inf.}\ }\textbf {\bibinfo {volume} {7}},\
  \bibinfo {pages} {152} (\bibinfo {year} {2021})}\BibitemShut {NoStop}%
\bibitem [{\citenamefont {Joshi}\ \emph {et~al.}(2020)\citenamefont {Joshi},
  \citenamefont {Elben}, \citenamefont {Vermersch}, \citenamefont {Brydges},
  \citenamefont {Maier}, \citenamefont {Zoller}, \citenamefont {Blatt},\ and\
  \citenamefont {Roos}}]{Joshi2020a}%
  \BibitemOpen
  \bibfield  {author} {\bibinfo {author} {\bibfnamefont {M.~K.}\ \bibnamefont
  {Joshi}}, \bibinfo {author} {\bibfnamefont {A.}~\bibnamefont {Elben}},
  \bibinfo {author} {\bibfnamefont {B.}~\bibnamefont {Vermersch}}, \bibinfo
  {author} {\bibfnamefont {T.}~\bibnamefont {Brydges}}, \bibinfo {author}
  {\bibfnamefont {C.}~\bibnamefont {Maier}}, \bibinfo {author} {\bibfnamefont
  {P.}~\bibnamefont {Zoller}}, \bibinfo {author} {\bibfnamefont
  {R.}~\bibnamefont {Blatt}}, \ and\ \bibinfo {author} {\bibfnamefont {C.~F.}\
  \bibnamefont {Roos}},\ }\href {\doibase 10.1103/PhysRevLett.124.240505}
  {\bibfield  {journal} {\bibinfo  {journal} {Phys. Rev. Lett.}\ }\textbf
  {\bibinfo {volume} {124}},\ \bibinfo {pages} {240505} (\bibinfo {year}
  {2020})}\BibitemShut {NoStop}%
\bibitem [{\citenamefont {Joshi}\ \emph {et~al.}(2022)\citenamefont {Joshi},
  \citenamefont {Elben}, \citenamefont {Vikram}, \citenamefont {Vermersch},
  \citenamefont {Galitski},\ and\ \citenamefont {Zoller}}]{LKJ_2022_chaos}%
  \BibitemOpen
  \bibfield  {author} {\bibinfo {author} {\bibfnamefont {L.~K.}\ \bibnamefont
  {Joshi}}, \bibinfo {author} {\bibfnamefont {A.}~\bibnamefont {Elben}},
  \bibinfo {author} {\bibfnamefont {A.}~\bibnamefont {Vikram}}, \bibinfo
  {author} {\bibfnamefont {B.}~\bibnamefont {Vermersch}}, \bibinfo {author}
  {\bibfnamefont {V.}~\bibnamefont {Galitski}}, \ and\ \bibinfo {author}
  {\bibfnamefont {P.}~\bibnamefont {Zoller}},\ }\href {\doibase
  10.1103/PhysRevX.12.011018} {\bibfield  {journal} {\bibinfo  {journal} {Phys.
  Rev. X}\ }\textbf {\bibinfo {volume} {12}},\ \bibinfo {pages} {011018}
  (\bibinfo {year} {2022})}\BibitemShut {NoStop}%
\bibitem [{\citenamefont {Dong}\ \emph {et~al.}(2024)\citenamefont {Dong},
  \citenamefont {Zhang}, \citenamefont {Dag}, \citenamefont {Gao},
  \citenamefont {Wang}, \citenamefont {Deng}, \citenamefont {Zhang},
  \citenamefont {Chen}, \citenamefont {Xu}, \citenamefont {Wang} \emph
  {et~al.}}]{dong2024measuring}%
  \BibitemOpen
  \bibfield  {author} {\bibinfo {author} {\bibfnamefont {H.}~\bibnamefont
  {Dong}}, \bibinfo {author} {\bibfnamefont {P.}~\bibnamefont {Zhang}},
  \bibinfo {author} {\bibfnamefont {C.~B.}\ \bibnamefont {Dag}}, \bibinfo
  {author} {\bibfnamefont {Y.}~\bibnamefont {Gao}}, \bibinfo {author}
  {\bibfnamefont {N.}~\bibnamefont {Wang}}, \bibinfo {author} {\bibfnamefont
  {J.}~\bibnamefont {Deng}}, \bibinfo {author} {\bibfnamefont {X.}~\bibnamefont
  {Zhang}}, \bibinfo {author} {\bibfnamefont {J.}~\bibnamefont {Chen}},
  \bibinfo {author} {\bibfnamefont {S.}~\bibnamefont {Xu}}, \bibinfo {author}
  {\bibfnamefont {K.}~\bibnamefont {Wang}},  \emph {et~al.},\ }\href
  {https://arxiv.org/abs/2403.16935} {\bibfield  {journal} {\bibinfo  {journal}
  {arXiv preprint arXiv:2403.16935}\ } (\bibinfo {year} {2024})}\BibitemShut
  {NoStop}%
\bibitem [{Note1()}]{Note1}%
  \BibitemOpen
  \bibinfo {note} {Note that we have defined $S_A^z$ and $S_A^x$ as twice the
  usual magnetizations in order that their eigenvalues are
  integers.}\BibitemShut {Stop}%
\bibitem [{Note2()}]{Note2}%
  \BibitemOpen
  \bibinfo {note} {The decoherences and their rates have been taken from
  Ref.~\cite {Brydges2019} and \cite {joshi-24}}\BibitemShut {NoStop}%
\bibitem [{\citenamefont {Andersen}\ \emph {et~al.}(2024)\citenamefont
  {Andersen}, \citenamefont {Astrakhantsev}, \citenamefont {Karamlou},
  \citenamefont {Berndtsson}, \citenamefont {Motruk}, \citenamefont {Szasz},
  \citenamefont {Gross}, \citenamefont {Westerhout}, \citenamefont {Zhang},
  \citenamefont {Forati} \emph {et~al.}}]{andersen2024thermalization}%
  \BibitemOpen
  \bibfield  {author} {\bibinfo {author} {\bibfnamefont {T.~I.}\ \bibnamefont
  {Andersen}}, \bibinfo {author} {\bibfnamefont {N.}~\bibnamefont
  {Astrakhantsev}}, \bibinfo {author} {\bibfnamefont {A.}~\bibnamefont
  {Karamlou}}, \bibinfo {author} {\bibfnamefont {J.}~\bibnamefont
  {Berndtsson}}, \bibinfo {author} {\bibfnamefont {J.}~\bibnamefont {Motruk}},
  \bibinfo {author} {\bibfnamefont {A.}~\bibnamefont {Szasz}}, \bibinfo
  {author} {\bibfnamefont {J.~A.}\ \bibnamefont {Gross}}, \bibinfo {author}
  {\bibfnamefont {T.}~\bibnamefont {Westerhout}}, \bibinfo {author}
  {\bibfnamefont {Y.}~\bibnamefont {Zhang}}, \bibinfo {author} {\bibfnamefont
  {E.}~\bibnamefont {Forati}},  \emph {et~al.},\ }\href
  {https://arxiv.org/abs/2405.17385} {\bibfield  {journal} {\bibinfo  {journal}
  {arXiv preprint arXiv:2405.17385}\ } (\bibinfo {year} {2024})}\BibitemShut
  {NoStop}%
\bibitem [{\citenamefont {Wang}\ \emph {et~al.}(2024)\citenamefont {Wang},
  \citenamefont {Zhou}, \citenamefont {Zhou},\ and\ \citenamefont
  {Zhang}}]{PhysRevLett.133.083402}%
  \BibitemOpen
  \bibfield  {author} {\bibinfo {author} {\bibfnamefont {C.-Y.}\ \bibnamefont
  {Wang}}, \bibinfo {author} {\bibfnamefont {T.-G.}\ \bibnamefont {Zhou}},
  \bibinfo {author} {\bibfnamefont {Y.-N.}\ \bibnamefont {Zhou}}, \ and\
  \bibinfo {author} {\bibfnamefont {P.}~\bibnamefont {Zhang}},\ }\href
  {\doibase 10.1103/PhysRevLett.133.083402} {\bibfield  {journal} {\bibinfo
  {journal} {Phys. Rev. Lett.}\ }\textbf {\bibinfo {volume} {133}},\ \bibinfo
  {pages} {083402} (\bibinfo {year} {2024})}\BibitemShut {NoStop}%
\bibitem [{\citenamefont {Zang}\ \emph {et~al.}(2024)\citenamefont {Zang},
  \citenamefont {Gu},\ and\ \citenamefont {Jiang}}]{zang-24}%
  \BibitemOpen
  \bibfield  {author} {\bibinfo {author} {\bibfnamefont {Y.}~\bibnamefont
  {Zang}}, \bibinfo {author} {\bibfnamefont {Y.}~\bibnamefont {Gu}}, \ and\
  \bibinfo {author} {\bibfnamefont {S.}~\bibnamefont {Jiang}},\ }\href
  {https://doi.org/10.1103/PhysRevLett.133.106503} {\bibfield  {journal}
  {\bibinfo  {journal} {Phys. Rev. Lett.}\ }\textbf {\bibinfo {volume} {133}},\
  \bibinfo {pages} {106503} (\bibinfo {year} {2024})}\BibitemShut {NoStop}%
\bibitem [{\citenamefont {Mao}\ \emph {et~al.}(2024)\citenamefont {Mao},
  \citenamefont {Zhai},\ and\ \citenamefont {Yang}}]{mao-24}%
  \BibitemOpen
  \bibfield  {author} {\bibinfo {author} {\bibfnamefont {L.}~\bibnamefont
  {Mao}}, \bibinfo {author} {\bibfnamefont {H.}~\bibnamefont {Zhai}}, \ and\
  \bibinfo {author} {\bibfnamefont {F.}~\bibnamefont {Yang}},\ }\href
  {https://doi.org/10.48550/arXiv.2402.15964} {\bibfield  {journal} {\bibinfo
  {journal} {arXiv:2402.15964}\ } (\bibinfo {year} {2024})}\BibitemShut
  {NoStop}%
\bibitem [{\citenamefont {McCulloch}\ \emph {et~al.}(2023)\citenamefont
  {McCulloch}, \citenamefont {De~Nardis}, \citenamefont {Gopalakrishnan},\ and\
  \citenamefont {Vasseur}}]{McCulloch2023}%
  \BibitemOpen
  \bibfield  {author} {\bibinfo {author} {\bibfnamefont {E.}~\bibnamefont
  {McCulloch}}, \bibinfo {author} {\bibfnamefont {J.}~\bibnamefont
  {De~Nardis}}, \bibinfo {author} {\bibfnamefont {S.}~\bibnamefont
  {Gopalakrishnan}}, \ and\ \bibinfo {author} {\bibfnamefont {R.}~\bibnamefont
  {Vasseur}},\ }\href {\doibase 10.1103/PhysRevLett.131.210402} {\bibfield
  {journal} {\bibinfo  {journal} {Phys. Rev. Lett.}\ }\textbf {\bibinfo
  {volume} {131}},\ \bibinfo {pages} {210402} (\bibinfo {year}
  {2023})}\BibitemShut {NoStop}%
\bibitem [{\citenamefont {Mezzadri}(2007)}]{Mezzadri2006}%
  \BibitemOpen
  \bibfield  {author} {\bibinfo {author} {\bibfnamefont {F.}~\bibnamefont
  {Mezzadri}},\ }\href {http://arxiv.org/abs/math-ph/0609050} {\bibfield
  {journal} {\bibinfo  {journal} {Notices Am. Math. Soc.}\ }\textbf {\bibinfo
  {volume} {54}},\ \bibinfo {pages} {592} (\bibinfo {year} {2007})}\BibitemShut
  {NoStop}%
\end{thebibliography}%
\end{document}